\documentclass[a4paper,UKenglish]{lipics-v2021}
\hideLIPIcs  %uncomment to remove references to LIPIcs series (logo, DOI, ...), e.g. when preparing a pre-final version to be uploaded to arXiv or another public repository

\usepackage{hyperref}
\hypersetup{%
      colorlinks    = true,%  Color text instead of boxes
  linkcolor = {blue},
   anchorcolor = {blue},
   citecolor = {blue},
   filecolor = {blue},
   urlcolor = {blue},
   menucolor = {blue},
   pagecolor = {blue},
}
\usepackage{xcolor}
\definecolor{blue}{rgb}{0.25, 0.41, 0.88}
\definecolor{blue}{rgb}{0.0, 0.22, 0.66}
\usepackage{booktabs}
\usepackage{siunitx}
\usepackage{etoolbox}
\usepackage{multirow}
\usepackage{xspace}
\usepackage{balance}
\usepackage[font=rm]{caption}
\usepackage{algorithm}
\usepackage[noend]{algpseudocode}
\usepackage{verbatim}
\usepackage{graphicx}
\usepackage{tabularx}
\usepackage{lscape}
\usepackage{shortvrb}
\usepackage{amsmath,amsfonts,amssymb,amsthm}
\usepackage{tikz}
\usepackage{soul}
\usepackage[all]{nowidow}
\usetikzlibrary{matrix,decorations.pathreplacing,calc,positioning,automata,arrows,fit,shapes.geometric,backgrounds}

\robustify\bfseries

\newcommand{\method}[1]{{\small\sf{#1}}}

\newcommand{\var}[1]{\mbox{\emph{#1}}}

\newcommand{\myparagraph}[1]{\paragraph*{#1}}

\newcommand{\maxscore}{\method{MaxScore}}

\newcommand{\aftertabspace}{\vspace*{2ex}}

\newcommand{\vbyte}{{\method{VByte}}}
\newcommand{\dvbyte}{{\method{Double-VByte}}}

\newcommand{\wand}{{\method{WAND}}}

\newcommand{\ddt}{{\ensuremath{d_{t,i}}}}
\newcommand{\gdt}{{\ensuremath{g_{t,i}}}}
\newcommand{\fdt}{{\ensuremath{f_{t,i}}}}
\newcommand{\wdt}{{\ensuremath{w_{t,i,j}}}}
\newcommand{\dtdf}{{\ensuremath{\langle \ddt, \fdt\rangle}}}

\newcommand{\dtdw}{{\ensuremath{\langle \ddt, \wdt\rangle}}}
\newcommand{\Index}{\mbox{$\mathcal{I}$}}

\newcommand{\Const}{\method{Const}}
\newcommand{\Expon}{\method{Expon}}
\newcommand{\Triangle}{\method{Triangle}}

\newcommand{\wsj}{\method{WSJ1}}
\newcommand{\robust}{\method{Robust04}}
\newcommand{\wiki}{\method{Wikipedia}}

\usepackage[square,numbers,sort]{natbib} 

\author{Alistair Moffat}{The University of Melbourne, Australia}{ammoffat@unimelb.edu.au}{}{}
\author{Joel Mackenzie}{The University of Queensland, Australia}{joel.mackenzie@uq.edu.au}{}{} %last two are orcid and funding

\authorrunning{A. Moffat and J. Mackenzie} 
\Copyright{A. Moffat and J. Mackenzie}

%{\textcolor{red}{Replace ccsdesc macro with valid one}} %TODO mandatory: Please choose ACM 2012 classifications from https://dl.acm.org/ccs/ccs_flat.cfm 

 %TODO mandatory; please add comma-separated list of keywords

\category{Author Version. Please refer to the published version if you plan to cite this work: {\url{https://doi.org/10.1016/j.ipm.2022.103248}}} %optional, e.g. invited paper

\relatedversion{}
\acknowledgements{}

\nolinenumbers %uncomment to disable line numbering

\title{Efficient Immediate-Access Dynamic Indexing}

\begin{document}

\maketitle

\begin{abstract}
In a dynamic retrieval system, documents must be
ingested as they arrive, and be immediately findable by queries.
Our purpose in this paper is to describe an index structure and
processing regime that accommodates that requirement for immediate
access, seeking to make the ingestion process as streamlined as
possible, while at the same time seeking to make the growing index as
small as possible, and seeking to make term-based querying via the
index as efficient as possible.
We describe a new compression operation and a novel approach to
extensible lists which together facilitate that triple goal.
In particular, the structure we describe provides incremental
document-level indexing using as little as two bytes per posting and
only a small amount more for word-level indexing; provides fast
document insertion; supports immediate and continuous queryability;
provides support for fast conjunctive queries and
similarity score-based ranked queries; and
facilitates fast conversion of the dynamic index to a ``normal''
static compressed inverted index structure.
Measurement of our new mechanism confirms that in-memory dynamic
document-level indexes for collections into the gigabyte range can be
constructed at a rate of two gigabytes/minute using a typical server
architecture, that multi-term conjunctive Boolean queries can be
resolved in just a few milliseconds each on average even while new
documents are being concurrently ingested, and that the net memory
space required for all of the required data structures amounts to an
average of as little as two bytes per stored posting,
less than half the space required by the best previous mechanism.
\end{abstract}

\section{Introduction}
\label{sec-intro}

An inverted index is a key component of most information retrieval
(IR) systems.
The standard inverted index structure consists of a
{\emph{vocabulary}} that maps strings to numeric term identifiers and
also stores any required global information about the term (for
example, the number of documents that contain it one or more times);
plus a set of {\emph{postings lists}} that record, for each term $t$,
the set of documents that $t$ appears in.
Each postings list is a sequence of {\emph{postings}}
that allows any bag-of-words query $Q$ to be
resolved by accessing the $|Q|$ postings lists for the query terms,
and then using the locational and occurrence information that they
contain as input to a similarity computation.
{\citet{zm06-csurv}} and {\citet{bcc10-ir}} provide overviews of
inverted index-based text querying.
If the set of documents that is the target for such queries is fixed
and stable, the index can be computed in a pre-processing phase, and
then used during querying as a static resource.
In this case there are two desired attributes which are likely to be
in tension: the index should be as compact as possible, to minimize
the storage footprint; and queries should be able to be executed as
quickly as possible, to both minimize the computational footprint
{\cite{greenir, mm+16-ipm}} and to improve the user experience
{\cite{bc19-ipm, ba+17-tois}}.

\begin{figure}[h]
\centering
\includegraphics[width=0.35\textwidth]{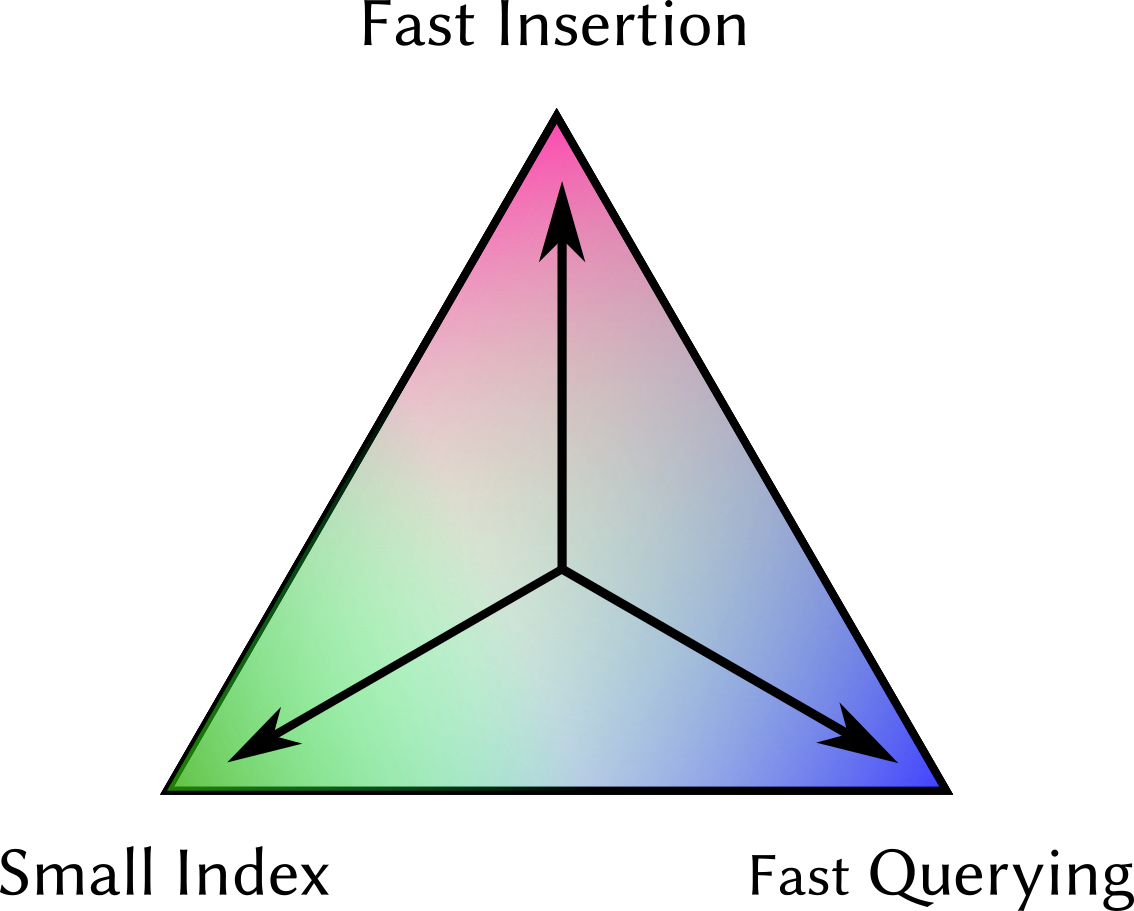}
  \caption{Trade-offs in the design of dynamic IR systems.
\label{fig-sausages}}
\end{figure}

On the other hand, if the set of documents is {\emph{dynamic}}, with
new documents arriving intermixed as part of a stream of operations
that includes insertions as well as queries, an additional
goal is introduced:
ingestion of documents should be as fast as possible, and should
be done in a manner that allows new documents to be immediately found
in response to subsequent queries.
That is, in a dynamic collection it is important that the index be
continuously queryable, even as documents are being added.
Figure~\ref{fig-sausages} illustrates the resultant three-way
operational tension, and thus the spectrum of
possibilities for designing extensible indexing and query 
processing systems.
Each of the three vertices represents a single
optimization goal, with the open space between them showing the
available span of three-way tradeoffs.
For example, while it is possible to achieve fast insertion by
regarding individual postings as elements in a linked list and
appending them one-by-one into a single array of nodes
{\citep{ewz22ecir}}, doing so is both expensive in terms of space,
and also costly in terms of query speed.
Similarly, compression techniques such as binary interpolative coding
are known to obtain highly compressed representations (for example,
see {\citet{pv21-compsurv}}), but they operate on whole postings
lists, meaning that incremental updates would require expensive
cycles of decompression and recompression.

\subsection{Goals and Contribution}

In this paper we provide new trade-off points in the space defined by
Figure~\ref{fig-sausages}.
In particular, we describe (Section~\ref{sec-somethingnew}) a
carefully engineered dynamic inverted index that balances the three
performance dimensions shown in Figure~\ref{fig-sausages} and
provides fast ingest of new documents, a
reduced storage footprint
compared to previous dynamic indexing techniques,
and efficient immediate-access querying.
Our approach is able to support immediate-access dynamic indexing on
typical document streams
such as Wall Street
Journal newspaper articles and Wikipedia articles using around two
bytes per posting in a document-level index to cover all index costs
(the vocabulary search structure, including the terms themselves and
their associated global term information, plus all of the postings
lists); and less than three bytes per posting if term-position
information is also required, for example, to support phrase or
proximity querying modes {\cite{cp08-ipm}}.
These compression rates mean that
the index requires less than half of the storage space of the most
recent dynamic immediate-access indexing method, that of
{\citet{ewz22ecir}}.
The economy of space that we have achieved in part arises because of
the innovative way in which we have structured the collection
vocabulary as a component of the first block of postings for each
term, and in part is a result of a novel byte packing operation that
allows substantial reduction in the average cost of storing the
postings.
The new byte packing operation also has applicability beyond text
indexing.
As a final innovation, the structure of our new immediate-access
index also allows for skip-links that allow fast conjunctive Boolean
querying to be performed.

Our experiments (Section~\ref{sec-experiments}) show
that our structure achieves an attractive balance of attributes, and
that we compare very favorably with the dynamic index scheme of
{\citet{hb17adcs}}, which offers similar ingest speeds, but builds
indexes that are more than twice as large.
Section~\ref{sec-theresmore} then considers a number of related
issues, including word-level indexing, and algorithmic issues
associated with extensible lists.
In particular, we take a fresh look at the question of how many
storage slots to allocate in response to an incremental request when
the final total size required is not (and may never be) known.
We show that a novel approach based on the triangle number sequence
(rather than the previously-employed arithmetic or geometric
sequences) has asymptotically smaller overhead
wastage, and generates additional memory space savings.

Measurement of the new mechanism confirms that in-memory dynamic
document-level indexes for collections into the gigabyte range can be
constructed at a rate of two gigabytes/minute using a typical server
architecture.
Further, our experiments also demonstrate that multi-term conjunctive
Boolean queries can be resolved in just a few milliseconds each on
average even while new documents are being ingested as part of the
same stream of operations, and that the net memory space required for
all of the required data structures amounts to an average of as
little as two bytes per stored posting.
 \section{Background}
\label{sec-background}

In this section we introduce a range of background material,
including previous proposals for dynamic index structures, thereby
setting the scene for the detailed development in
Section~\ref{sec-somethingnew}.

\subsection{Inverted Indexing}
\label{sec-indexing}

In a {\emph{document-level}} inverted index each postings list
contains postings of the form $\dtdf$, with $\ddt$ the ordinal
identifier of the $i$\,th document containing term $t$, and $\fdt$
the corresponding within-document frequency that counts the number of
occurrences of $t$ in $\ddt$.
It is usual to store the postings in {\emph{document order}}, that
is, sorted by increasing values of $\ddt$, and for the original set
of ordinal document numbers to be converted to a sequence of
{\emph{$d$-gaps}}, $\gdt$, via the transformation $g_{t,1}=d_{t,1}$,
and thereafter $\gdt=\ddt-d_{t,i-1}$.
Conversion to gaps renders the postings lists more compressible (see
Section~\ref{sec-compression}), and we employ gap-based document
numbers throughout this paper.
In particular, even when we refer to a $\dtdf$ tuple, what is
actually stored is a $\langle\gdt,\fdt\rangle$ tuple.
Note that $\gdt\ge1$ in all postings in a document-level index, since
there is at most one posting per term per source document.
{\citet{zm06-csurv}} and {\citet{bcc10-ir}} further explain inverted
indexing and provide examples.

\begin{table}
\centering
\begin{tabular}{ccl}
\toprule
Type & Postings & Explanation
\\
\midrule
document-level
	& $\dtdf$
		& $d$-gaps $\gdt>0$ and corresponding frequencies $\fdt>0$
\\
word-level
	& $\langle \ddt, \fdt, \hat{w}\rangle$
		& as above, and adding a list of $w$-gaps for word positions
\\
word-level
	& $\dtdw$
		& $d$-gaps $\gdt\ge0$ and $w$-gaps
			$w_{t,i,j}>0$ within document $\ddt$
				
\\
word-level
	& $\langle w_{t,j}\rangle$
		& $w$-gaps $w_{t,j}>0$, without regard to document boundaries
\\
\bottomrule
\end{tabular}\aftertabspace
 \caption{Four different types of inverted index, in each case for
some term $t$, with $\ddt$ the $i$\,th document that contains $t$,
and with $\fdt$ the number of instances of $t$ in $\ddt$.
\label{tbl-indexes}}
\end{table}

In a {\emph{word-level}} inverted index the location of each word in
each document is also tracked.
One way this can be done is via postings that include a third
component $\hat{w}$ that contains a list of the $\fdt$ ordinal word
positions of $t$ within that $\ddt$\,th document, themselves stored
as a list of $w$-gaps relative to the start of that document.
A second approach -- and the option employed by {\citet{hb17adcs}},
and also the one we make use of in our work here -- is to store
tuples $\dtdw$, where $\wdt$ is the word position of the $j$\,th
instance of term $t$ within document $\ddt$.
The $\wdt$ sequence across consecutive postings for any given
document $\ddt$ can again be reduced to $w$-gaps.
However in this representation care must be taken with the $d$-gaps
that form the first component of each posting, because now there
might be multiple postings for $t$ within a document $d$, meaning
that $\gdt\ge0$ is the best that can be assured.

The third option for word-level indexing completely separates the $d$
components from the $w$ components, with each postings list simply a
list of ordinal term positions in the entire collection, without
explicit inclusion of document numbers.
The resulting postings sequence $\langle w_{t,j}\rangle$, in which
$w_{t,j}$ is the word position in the collection of the $j$\,th
occurrence of term $t$, can be stored as a list of
whole-of-collection $w$-gaps, to reduce the space required; this is
the approach employed by {\citet{bc05tr}}.
The drawback of this third option -- which is sometimes referred to
as being a {\emph{schema-independent index}} -- is that when
corresponding document numbers are required, they must be determined
via binary (or other) search in an array that records, for each
ordinal document number, its first word number in the collection.
On the other hand, the schema-independent approach is the most
compact of the three -- placing it at a different location in the
space of options described by Figure~\ref{fig-sausages}.

Table~\ref{tbl-indexes} summarizes the properties of these four types
of inverted index.
Our new implementation of immediate-access indexing explores the
first and the third of the four possibilities shown.
The choice between the three word-level options was in part
determined by typical querying use cases, including the need for
document boundaries to be distinguishable; and in part by the
desirability of each posting containing two components, the benefit
of which is explained in Section~\ref{sec-magic}.

\subsection{Index Compression}
\label{sec-compression}

Compact storage of the postings list is essential, since they
comprise the great bulk of any index, especially word-level ones.
Fortunately, once ordinal values have been reduced to $d$- and/or
$w$-gaps, a range of highly effective integer compression techniques
can be applied, see {\citet{pv21-compsurv}} for a survey.
In conjunction with document reordering techniques
{\citep{ao+18-ipm,dk+16-kdd,ieeetkde22mpm}}, the best of those
integer coding techniques can typically represent document-level
postings in less than eight bits each on
average, a measurement we refer to as
being the {\emph{compression effectiveness}}.
However, both document reordering and high-effectiveness integer
coding regimes are based upon holistic analysis of the gapped values
making up each posting list, an option that is not available in
dynamic indexing applications.

Instead, dynamic indexes usually make use of {\emph{byte codes}},
collectively referred to as being the {\vbyte} mechanism (although
there are also small differences between specific descriptions).
This form of code has been known for more than five decades.
For example, {\citet{heaps72infor}} describes a general approach to
integer compression that includes arrangements in which the code
lengths are $8$, $16$, $24$ (and so on) bits long; and
{\citet{cp90sigir}} also describe a {\vbyte} mechanism.
{\citet{wz99compjour}} include {\vbyte} in their experimental study,
and further comparison was undertaken by {\citet{swyz02sigir}} and by
{\citet{tro03irj,t14-adcs}}.
Subsequent developments are then reported by {\citet{mnzb00tois}}, by
{\citet{bfne03spire}}, by {\citet{cm05spire}}, by {\citet{d09-wsdm}},
by {\citet{sg+11-cikm}}, and most recently, by {\citet{lkr18-ipl}}.

Encoding non-negative integers $x\ge0$ using {\vbyte} is
straightforward, and operates quickly.
At each iteration the low-order seven bits of $x$ are isolated via a
mask, and the remaining high-order bits are considered.
If those high-order bits are zero, then the low-order bits are
written as a byte with a 1-bit in the vacant top position, the latter
indicating that $x$ is now finished.
On the other hand, if the high-order bits of $x$ are non-zero, the
low-order seven bits are written as a byte value between $0$ and
$127$, and the process is iteratively applied to the value $x'$ that
contains the high-order bits of the original value $x$, shifted seven
bits right.
That is, each seven-bit segment of $x$ is written within an eight-bit
byte, with a ``{\tt{0}}'' top-bit indicating ``continue, another byte
is required'', and a ``{\tt{1}}'' top-bit indicating ``stop now, this
is the last byte for this value''.
For example, the decimal number $12{,}345$ requires fourteen bits in
binary, ``{\tt{1100000\,0111001}}'', and spans two seven-bit
segments.
It would thus be written as two eight-bit bytes:
``{\tt{{\color{gray}{0}}\,0111001}} +
{\tt{{\color{gray}{1}}\,1100000}}'', where the leading bits in each
byte are the flag bits that indicate whether to continue or to stop.
Decoding using {\vbyte} methods is also fast, involving mask and
shift operations only.

{\citet{bc05tr}} give pseudo-code for the {\vbyte} variant that we
employ here.
One key point to note in connection with this version is that there
is only one possible cause of an all-zero byte in the compressed
sequences -- the pattern ``{\tt{00000000}}'' (a null byte) can only
arise as the code for an input value $x=0$.
This fact means that if we ensure that $x>0$ at all encoding steps,
then null bytes will never appear as part of a coded integer,
allowing ``{\tt{00000000}}'' to be used as a sentinel to mark the end
of a sequence of compressed values, without the length of the
sequence (nor its compressed length) needing to be stored explicitly.
When a {\vbyte} decoder encounters a null byte, the compressed
sequence has ended.

While {\vbyte} is a static code and thus needs no pre-analysis of the
data that it is to be applied to, for document-level postings list
compression it has the disadvantage of requiring a minimum of two
bytes per posting, one for the $\ddt$ value and one for the $\fdt$
component (or for the $\wdt$ component in a word-level index stored
as $\dtdw$ postings).
Section~\ref{sec-magic} introduces a packing protocol that addresses
that drawback and reduces the average cost of $\dtdf$ postings for
typical document collections to approximately $1.5$ bytes per
posting, allowing a substantial space reduction to be achieved.

Many other highly-effective compression methods exist.
But they typically act holisticly on blocks of postings
{\cite{lb15-spe,ms00-irj,pv21-compsurv,milc}}, thereby creating a
tension between storage space and updateability, as has already been
noted in the trade-off spectrum shown in Figure~\ref{fig-sausages}.
Static compression techniques are the methods of choice for
immediate-access indexing.

\subsection{Static Index Construction}

Early index construction techniques were designed for static
collections, and operated as two or more processing phases, often
also relying on external disk storage.
For example, {\citet{fl91tr}}
describe a multi-pass arrangement in which each traversal of the
input text builds the postings lists for a subset of the terms, with
the number of passes dictated by the relationship between the total
number of postings in each fragment of the index, and the amount of
main memory available to store them while the lists are constructed.

{\citet{hc90jasis}} describe an in-memory inversion process, using a
binary search tree to store the vocabulary of the collection, with a
linked list of postings attached to each tree node to thread together
the term observations.
At the end of the input the index is recovered by traversing the tree
in term order, and at each node, following the list links.
In a different approach, {\citet{mof92compsys}} showed that for one
type of efficient postings list compression method the cost of
storing the postings lists in compressed form could be upper-bounded
based solely upon knowledge of the term frequencies, and described a
two-pass inversion process that first builds a collection vocabulary
and counts term frequencies; then allocates the starting points of
the postings lists of a compressed inverted index in memory; and
then, in a second pass, computes the $d$-gaps and $\fdt$ values, and
fills in the compressed index in a random-access order.
While still limited by the amount of main memory available, the use
of effective compression meant that relatively large amounts of text
could be efficiently handled.

In subsequent work {\citet{mb95jasis}} remove the limit imposed by
main memory, while still using compression as an important part of
the process.
In this approach incoming documents are processed in batches to
accumulate sets of posting that fit within the available memory
limit, and when that limit is reached, the postings are sorted and a
local index for that document batch is written to disk in compressed
form.
When all of the documents have been ingested, the set of partial
indexes is combined via a multi-way sequential in-place on-disk merge
operation that maintains -- and wherever possible, improves --
compression effectiveness, so as to operate within the same envelope
of disk space as was consumed by the set of compressed partial
indexes.
The method of {\citeauthor{mb95jasis}} can thus be regarded as being
complementary to the earlier work of {\citet{fl91tr}} -- the latter
manage large collections by splitting the vocabulary and making
multiple passes, whereas {\citeauthor{mb95jasis}} recommend instead
that the collection be partitioned on a ``by documents'' basis.

{\citet{hz03jasist}} added further enhancements.
They observed that there is no requirement for a comprehensive
vocabulary covering all terms to be constructed until the end of the
inversion process, and that the vocabulary can thus also be created
in batches.
In their proposal batches of documents are again inverted using as
much of the available memory as can be made available, maximizing the
number of documents in each batch so as to stay within that memory
limit.
They also directly build localized posting lists using compression,
rather than accumulating raw postings for later sorting, increasing
the number of documents handled in each batch, and hence reducing the
number of batches required.
Once constructed, each partial (compressed) index is written to disk,
including its own local vocabulary.
Then, when all of the document batches have been processed, the set
of local vocabularies is merged to make a whole-of-collection
vocabulary; and the terms' postings lists are read, interleaved as
required, and then written.
Only one pass is made over the source document collection, and
although there is a second processing phase to carry out the merge,
it operates over data that has already been processed into binary
format, and is not as expensive as a full pass over the source text
would be.

\subsection{Query Processing Using Static Indexes}
\label{sec-pruning}

Static indexes naturally lend themselves to efficient querying
algorithms, with a wide range of such techniques having been
developed.
One of the keys for achieving efficiency over static inverted indexes
is to embed ``skip pointers'' into the index at the time it is
constructed {\citep{mz96-tois}}, so that runs of postings can be
quickly bypassed if they are not required by the current query, thus
saving both storage accesses and potentially costly decoding
operations.
Skip pointers are critical for efficiently supporting fundamental
querying modes such as Boolean conjunctions
{\citep{cm10-tois,k+18-vldb}}.

Disjunctive query modes, which return the top-$k$ documents based on
some similarity estimation calculation, also benefit from static
indexes.
By employing a combination of skip pointers, encoded document scores
within postings lists, and a query-time data structure to record the
best scores ``seen so far'', it is possible to bypass documents known
to have no prospect of being among the final top-$k$ results.
These approaches, known collectively as {\emph{dynamic pruning}},
have been a focus of attention through several decades
{\cite{bc+03-cikm,dm+17-irj,ds11-sigir,mo+17-sigir,tf95-ipm}}.
We refer the interested reader to the surveys of
{\citet{tmo18-fntir}} and {\citet{mm20cikm}} for further details of
these.

\subsection{Dynamic Indexing and Querying}
\label{sec-dynamic}

Each in-memory index assembled during the operation of the
{\citet{hz03jasist}} approach is ostensibly one step towards a global
index for a large document collection.
But at any given instant the in-memory component can be configured to
provide the functionality of an operational inverted index for the
current batch of documents, and any queries that might arise can be
processed against the current document batch in the normal manner.

As part of their index, {\citet{hz03jasist}} maintain postings lists
using {\emph{extensible arrays}}, one per term.
Extensible arrays are a well-known data structure: when an array of
unknown eventual size is required, it is allocated $n_1=1$ cells at
first; thereafter, at each subsequent insertion, if the array is full
and all $n_z$ cells are occupied, a new array of size $n_{z+1}=\lceil
k\cdot n_z\rceil$ is allocated for some $k>1$ (with $k=2$ being a
typical value); the contents of the old array are copied to the new
one; the space associated with the old array is released; and then
the process is continued, now with $n_{z+1}-n_z-1$ empty slots in the
new array (after the one that triggered the expansion got claimed).
The geometric sizes determined by the multiplicative growth rate $k$
mean that when the array has reached $n$ elements the total number of
item copies over all of the completed growth cycles is bounded above
by $nk/(k-1)$; that is, the total time required to sequentially
append $n$ elements to an initially empty array is $O(n)$.
However extensible arrays are not so efficient in terms of space.
At each expansion moment an extensible array of size $n_z$ is fully
occupied, and another array of size $n_{z+1}$ has been allocated and
is awaiting its contents, at a time when $n=n_z+1$.
That is, there are repeated instants at which a total of
$(2+k)(n-1)\ge 2n$ cells are allocated to the storage of $n$
elements.

{\citet{bc05tr}} adopt much of the approach of
{\citeauthor{hz03jasist}}, but with one critical difference -- they
make use of {\emph{extensible lists}}, in which blocks of items are
linked together using pointers, with each postings list stored as a
chain of such blocks.
They argue that the overhead associated with maintaining one pointer
per block, plus having a small amount of space unused in the last
block, is a more attractive compromise than the
multiplicatively-growing unused space incurred by an extensible
array.
They further argue that during querying the need to follow a pointer
from time to time is an acceptable overhead.

{\citeauthor{bc05tr}} then go on to consider two strategies for
setting the block size: in the $\Const_B$ approach, each block is $B$
words, with one word required for the pointer to the next block, and
$B-1$ words available for {\emph{payloads}}, that is,
$B_{z+1}=B_z=B$.
In the alternative $\Expon_{B,k}$ approach, the payload capacities
grow as a geometric sequence, $B_{z+1}=1+\lceil k\cdot
(B_z-1)\rceil$, with $B_1=B$, and the assumption again being that
each pointer requires the same space as one payload element.
At any given moment a chain of $z$ blocks thus contains up to
$n=\sum_{i=1}^{z} (B_i-1)$ payloads, and $z$ pointers.
{\citet{bc05tr}} experiment using both synthetic Zipfian data and
also two different TREC collections, and conclude that the
$\Expon_{B=16,k=1.1}$ method, modified by the addition of an upper
cap of $256$ on block size, provides the best balance in terms of
indexing memory and indexing throughput.
With that approach they are able to obtain $\langle
w_{t,j}\rangle$-format word-level indexes
(see Table~\ref{tbl-indexes}) that require around $1.8$ bytes per
posting on typical TREC document collections, not including the
vocabulary and term structure costs.

{\citet{alb12kdd}} and {\citet{bgllll12icde}} also consider the
challenge of dynamic indexing and immediate-access search, with an
emphasis on the requirements of the Twitter search service.
They note that tweets are often required in reverse chronological
order in response to Boolean searches, and hence employ uncompressed
postings stored as $32$-bit words, reverse-chaining segments of size
$2^1$, $2^4$, $2^7$, and $2^{11}$ postings.

In other recent work, {\citet{hb17adcs}} revisit the mechanisms proposed
by {\citet{bc05tr}}, exploring further blends of $\Const$ and
$\Expon$ extensible lists, and considering the way that those
strategies interact with the non-linear behaviors introduced by
specific hardware characteristics.
{\citeauthor{hb17adcs}} make only limited use of compression, and the
most compact index they obtain for a $\dtdw$-format word-level
inverted index against a collection of $8.46 \times 10^9$ web pages
requires around $6.2$ bytes per posting, again not including the
vocabulary and hash table costs.

\subsection{Other Related Work}

The in-memory techniques of {\citet{hz03jasist}}, {\citet{bc05tr}},
and {\citet{hb17adcs}} all yield indexes for a batch of documents.
Other work has investigated how best to join such partial indexes
together.
For example, {\citet{lzw06ipm}} propose that each index batch be
immediately merged with a single on-disk index.
This approach results in growing merging costs as the two
components of each merge become increasingly disparate in size, and
is not asymptotically efficient; but also has the advantage that any
given query requires only two collections to be searched, the main
on-disk index, and the current in-memory batch index.

In contrast, {\citet{lmz08acmtods}} (see also {\citet{bc05cikm}})
propose a hierarchical merging strategy that always joins components
of approximately equal size and is asymptotically efficient, but
requires that queries be processed against both the current in-memory
index and a logarithmic-bounded number of on-disk indexes, with the
base of the logarithm determined by the batch size possible.
{\citet{bc08irj}} consider a range of hybrid strategies that blend
these two approaches.
The work we describe in this paper can be used as a first processing
phase and combined with any of these memory-to-disk merging
strategies.

Earlier work also considered dynamic index construction and querying
operations.
For example, {\citet{cp90sigir}} consider the batching of updates
into an inverted index stored as records in a B-tree structure, and
show that it is better to use any extra memory to accumulate postings
into batches than it is to cache upper-level B-tree nodes; they also
consider the case when postings must be displaced from main memory on
to disk.
The work of {\citet{fg92vldb}} is also directly relevant here: they
consider extensible arrays, extensible lists, and also a hybrid that
combines them, in each case regarding the set of postings as the
leaves of a B-tree.
They also explore several of the growth strategies that have already
been mentioned, including $\Const$ and $\Expon$, analyzing their
relative performance against a Zipfian distribution of occurrences,
and measuring storage effectiveness (but not employing any form of
compression) and search times using datasets derived from the
$11{,}657$ first and family names extracted from a university phone
directory.

{\citet{stg94sigir}} and {\citet{tgs94sigmod}} also examine Zipfian
distributions, and suggest the use of two different types of postings
list elements, short ones that store only a small number of postings
each, but also reduce the risk of excessive unused space being
associated with ``tail'' terms; and long ones with a greater postings
capacity, to be allocated when sufficient evidence of demand for that
term has been observed.
As documents are received and postings accumulated, terms are
dynamically shifted from the short-block category to the long-block
category if they exceed their initial space allocation, or if other
terms arise that pressure them out of their current block.

{\citet{bcc94vldb}} also describe dynamic indexing as being a process
in which objects (postings lists) grow and must be extensible.
At any given moment each posting list is one of a small palette of
possible sizes ($16$, $32$, $64$, through to $8192$ bytes), with
long lists that exceed $8192$ bytes handled by chaining blocks
together, another hybrid scheme.
{\citet{ccb94tr}} also make use of batch update operations to extend
the on-disk index needed when the collection is too large for its
index to fit into main memory.

Extensible arrays and extensible lists incur ``tail wastage'' that
occurs whenever the last allocated block is only part full.
{\citet{sc05ipm}} seek to address that issue by monitoring the way in
which lists accumulate postings, and employ statistical prediction
techniques in connection with extensible arrays.
They demonstrate higher space utilization and reduced relocation
levels compared to the $\Expon_{2.0}$ and $\Expon_{1.5}$ mechanisms
used as baselines.
The new approach to extensible lists that we present in
Section~\ref{sec-triangle} has the benefit of reducing the tail
wastage to an arbitrarily small fraction of the stored content as
list sizes become larger, without needing to employ statistical
estimation mechanisms.

In more recent work, {\citet{ewz22ecir}} consider dynamic indexing
from a different perspective.
They address the question of {\emph{sliding window retrieval}} in
which recent documents must be queryable, but older documents are to
be ``forgotten''.
Their proposal, dubbed the {\emph{apoptosic index}}, makes use of a
circular array of $n$ elements to retain the most recent $n$
postings, with older postings expiring as new ones are received.
Each posting contains four items, $\langle d, t, f_{t,d}, p\rangle$,
where $p$ is the position in the circular buffer of the previous
instance of term $t$, and hence provides a back-pointer that can be
used as a linked list during querying to step through the postings
associated with $t$.
In this arrangement -- with careful attention paid to relative
pointer positions -- expired postings are simply overwritten, without
a specific deletion operator being required.
That is, postings are replaced on a one-for-one basis, meaning that
if a certain number of documents are to be retained in the index, a
maximum ratio of postings to documents must be known in advance.
{\citeauthor{ewz22ecir}} report experiments using an implementation
in which the four posting components are represented as integers,
with a set of $n$ postings requiring $16n$ bytes (not including the
cost of the vocabulary or term search structure), a non-trivial
requirement when $n$ is large.
{\citeauthor{ewz22ecir}} observe that compression can be applied to
their postings, claiming that ``$16$--$20$ bits per node [posting]
should be straightforward''.
But with each back-pointer spanning a distance of one or more
multiples of the average number of distinct terms per document (that
is, many hundreds of positions in the circular array), that statement
seems optimistic.
Nor can the apoptosic method be used as a stepping stone in the
construction of a conventional on-disk inverted index.

Finally in this section, we note the
possibility of the appended documents not being incrementally indexed
at all, and instead being searched in text format using sequential
pattern-matching approaches.
While this approach is easy to implement, the batch of inserted
documents must eventually be indexed if the overall system is to be
long-term efficient, and so deferred indexing is only plausible
in highly restricted circumstances, such as when each document batch
is to be filtered before indexing, with only a subset of them needing
to be permanently retained.
 \section{Compact Immediate-Access Indexing}
\label{sec-somethingnew}

We now present our proposed structure, supposing that a stream of
incoming documents is to be converted to a document-level inverted
index (see Table~\ref{tbl-indexes}; word-level indexes are discussed
in Section~\ref{sec-wordlevel}), and that any new document must be
instantly findable via a concurrent query stream.
The key elements to note in this section are the role
that immediate-access indexing plays within a larger system in which
some parts of the index have already been reorganized into highly
efficient static arrangements (Section~\ref{sec-mode}); our use of
fixed-length postings blocks, the first of which also includes all of
the vocabulary information (Section~\ref{sec-fixedblocks}); the
chaining together of the blocks for each term as documents are
inserted (Section~\ref{sec-addingdocs}); the development of a packed
{\dvbyte} compression mechanism that brings a dramatic reduction in
the average cost of compressed postings (Section~\ref{sec-magic});
and the preliminary compression and throughput
results showing how effective it is
(Section~\ref{sec-prelimresults}).

\subsection{Mode of Operation}
\label{sec-mode}

\begin{figure}
\centering
\includegraphics[width=0.75\textwidth]{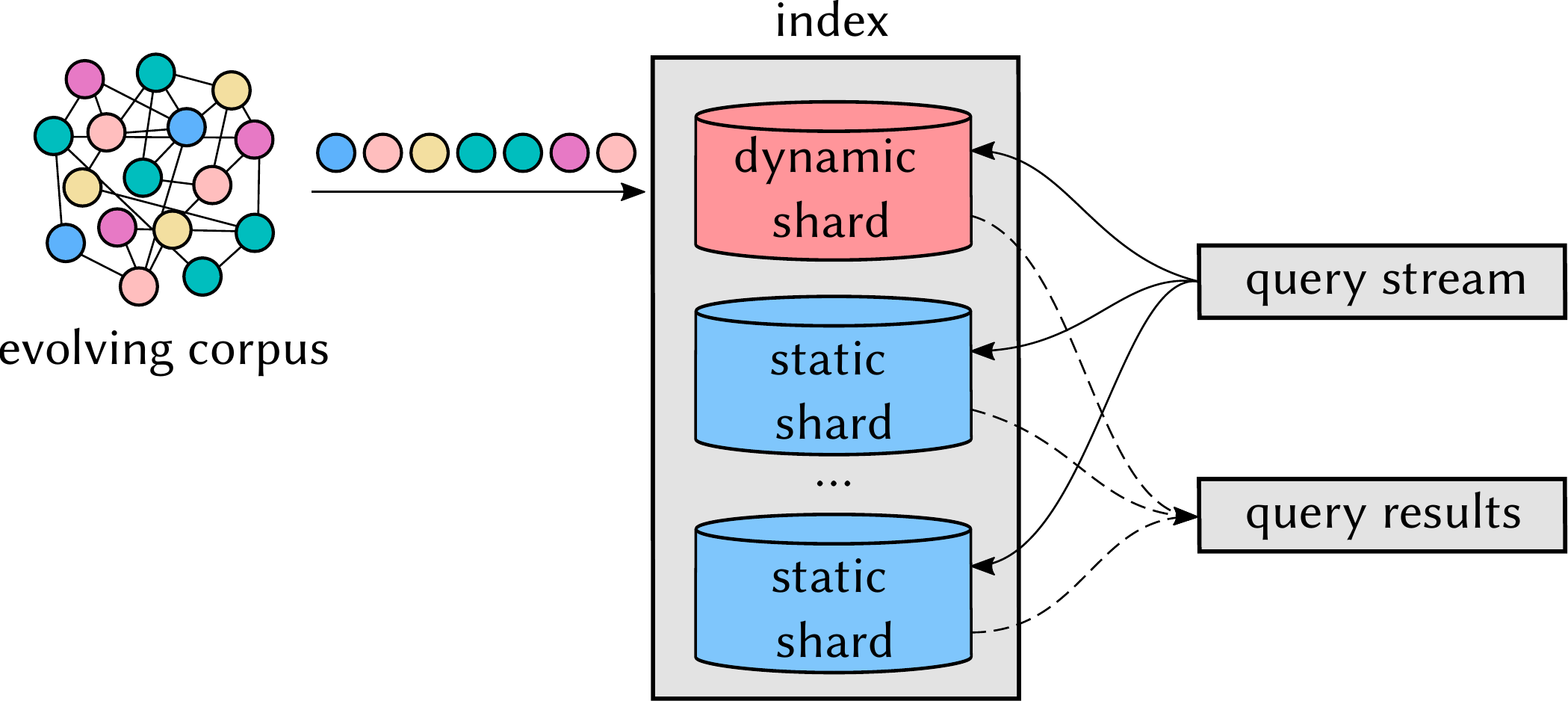}
\caption{Overall structure of a large-scale search
system in which documents are ingested and then permanently retained.
An in-memory dynamic shard index accumulates arriving documents, and
from time to time is shifted to secondary storage and reorganized to
form a further static shard index.
Each incoming query is processed against the dynamic shard index and
all previous static shard indexes, and the results fused.
The dynamic shard and static shards might have different index
organizations, and might employ different query processing
heuristics.
Our focus in this paper is on the dynamic shard index, and the
operations needed to ingest documents and to query against them.
\label{fig-mode}}
\end{figure}

Figure~\ref{fig-mode} illustrates the role that an immediate-access
dynamic index plays in a larger retrieval system.
Ingested documents are immediately added to the dynamic index and
rendered findable, a process that continues as long as the dynamic
shard index can be retained in main memory.
At the same time, previously accumulated document batches have had
their index data reorganized (and possibly merged) so as to maximize
retrieval speed and minimize stored space, thereby occupying a point
on the bottom edge in Figure~\ref{fig-sausages}; they can be searched
using the efficient querying techniques summarized in
Section~\ref{sec-pruning}.
When the current dynamic shard index has reached the available memory
limit it too is converted to the same static form, to make it smaller
and allow faster querying modes to be used.
After being reorganized it is added -- perhaps after being merged
with other static index components -- to the part of the index held
in secondary storage.
A new dynamic index is then initiated.

Because it faces different operational requirements,
the dynamic shard has a different internal index structure, and
occupies a different location in Figure~\ref{fig-sausages}.
Incoming queries are resolved against all shards, using the
corresponding strategy for the static and dynamic components, and
fused answer sets then returned to the users or passed to a
more sophisticated ranking mechanism.
Our focus in this paper is exclusively on the construction and
querying of the dynamic shard index, as a critical component of the
larger search ecosystem illustrated in Figure~\ref{fig-mode}.

Given this context, our research goals
can now be stated
\begin{itemize}
\item How should a dynamic index be organized so as to best balance
insertion cost, storage cost, and querying costs?
\item
To what extent do those necessary compromises erode compression and
retrieval performance relative to state-of-the art static indexing
and querying arrangements?
\end{itemize}
As a result of exploring those goals, and relative to
the background established by the various methods summarized in
Section~\ref{sec-background}, we are able to offer the following
points of distinction:
\begin{itemize}
\item
We describe an immediate-access indexing scheme that provides for
ingest of new documents at speeds comparable to those of
{\citet{hb17adcs}};
\item
At the same time, use of a new compression approach means that our
approach can process documents batches of roughly twice the size as
{\citet{hb17adcs}} within the same amount of memory; plus
\item
We document querying speeds in the immediate-access index, and while
these are (unsurprisingly) slower than can be attained in a
fully-optimized static retrieval system, they are neverthess fast
enough that the overall static/dynamic structure shown in
Figure~\ref{fig-mode} becomes practical.
\end{itemize}

\subsection{Fixed-Block Indexing}
\label{sec-fixedblocks}

As do {\citet{bc05tr}} and {\citet{hb17adcs}},
we make use of chains of blocks, one per indexed term, with each
block containing compressed postings.
But compared to those methods
there are also some key differences.
First is that the vocabulary information is stored as a component of
the first block associated with each term, rather than in a separate
data structure, to reduce the overheads associated with both the
variable-length term information and also with the first few
postings.
Packing these two variable-length entities together replaces and
improves upon the earlier suggestion of {\citet{tjc13adcs}} that one
or two postings be stored in the vocabulary.
Second, we assume
-- in this section, at least -- that all blocks are of the same
length (the $\Const$ strategy) and include an $h=4$-byte linking
pointer within a $B$-byte total size; but also consider a novel
variable-length block alternative in Section~\ref{sec-triangle}.
With fixed-size blocks rather than a sequence of
lengthening blocks, the index as a whole can thus be thought of as
an array of blocks, with array offsets used to locate blocks within
that array, rather than requiring byte-addressed pointers.
In turn, this cuts down on the overhead cost of the
meta-data needed to manage the index, and permits higher data
loadings to be achieved.
And third, we introduce (as an orthogonal
improvement) a new way of compressing postings that reduces the cost
of storing them by as much as one third compared to the usual
{\vbyte} mechanism.

In particular, by maintaining just one block size,
but employing around half of each term's first block to contain the
vocabulary information and leaving half (or even less) for the first
few postings, we in effect have two different block sizes in
operation: short first postings blocks, and thereafter
constant-length normal blocks, both containing exactly $B$ bytes in
total.
That is, in this new arrangement we
gain the same benefit as in previous approaches that used postings
blocks of increasing capacities, without the complexity of needing to
manage a range of memory allocation amounts.

\begin{figure*}[t]
\centering
\includegraphics[width=0.85\textwidth]{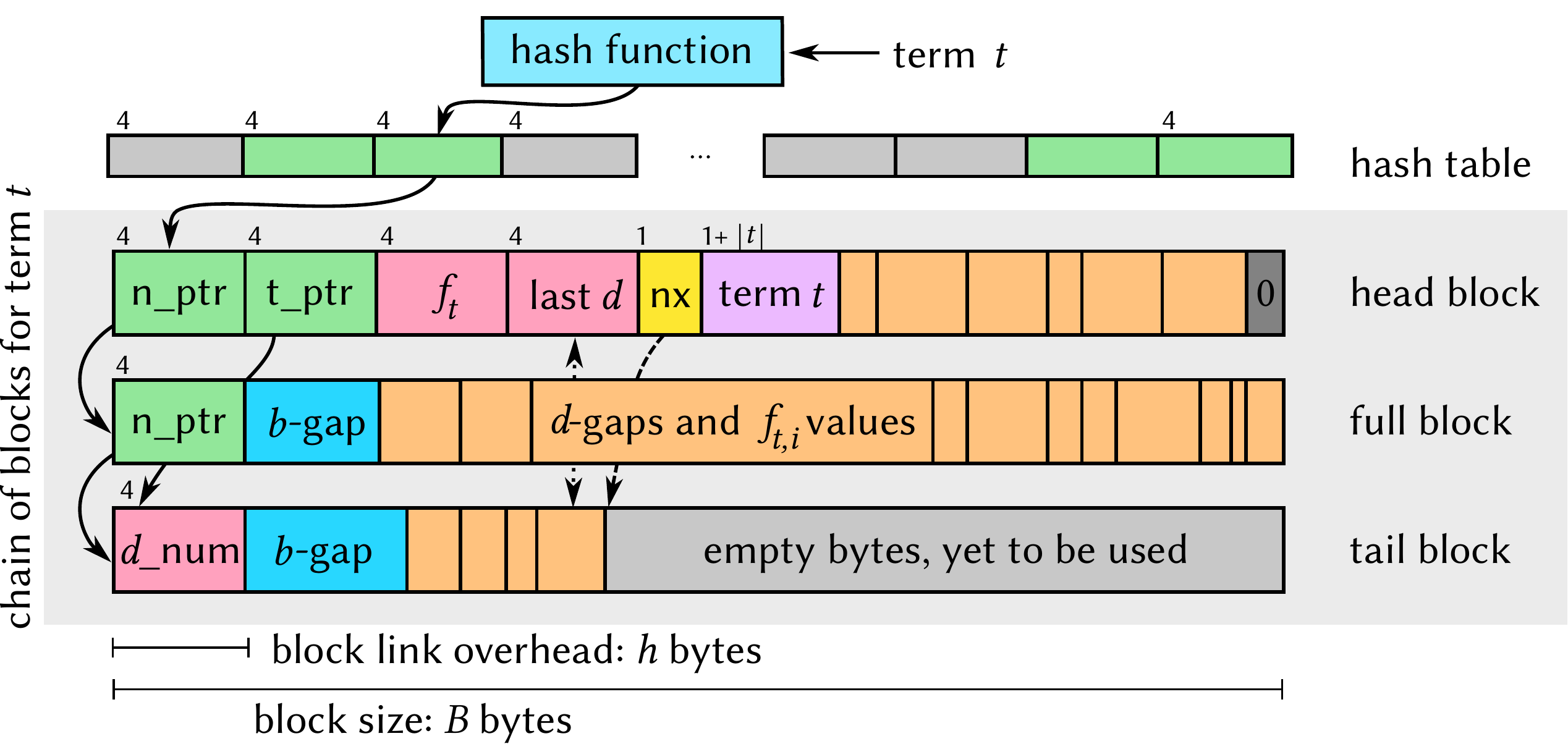}
\caption{The chain of $B$-byte blocks associated with one term $t$,
showing the three types of blocks that might arise and the way in
which the vocabulary (global whole-of-term information) and postings
(localized information) are distributed across them.
The superscript numbers besides some of the fields indicate the
length in bytes of those elements.
The $d$-gaps, $\fdt$ values, and $b$-gaps are all variable length
codes.
Any unused trailing bytes are assigned the value character null (as
shown at the end of the head block).
\label{fig-layout}}
\end{figure*}

Figure~\ref{fig-layout} illustrates these ideas, showing the chain of
blocks associated with one term $t$, noting the three types of blocks
that arise ({\emph{head}}, {\emph{full}}, and {\emph{tail}}) and
showing the fields associated with each.
The head block is located via a hash table, and contains all the
standard vocabulary information as well having the capacity to store
a modest number of postings.
Subsequent blocks of postings are then linked by pointers (denoted as
{\var{n\_ptr}} in the diagram), and by a single tail pointer (shown
as {\var{t\_ptr}}) that allows direct access from the head block to
the tail block.
The latter is normally only partially full, and is the site of any
update operations for this term,
with the field $\var{nx}$ in the head block indicating the offset in
the tail block of the current ``write'' location as that block gets
filled.
Term $t$'s blocks are interleaved with the blocks of other terms (not
shown in Figure~\ref{fig-layout}) in a single large array of
fixed-length blocks which we denote as $\Index$.
Each postings list contains one head block; zero or more full blocks;
and one tail block, which is initially also the head
block; overall, the index contains one head block for
each distinct term.

All blocks store postings $\dtdf$ coded using {\vbyte} (or as we
shall describe shortly, a more efficient variant of it).
In addition, the first document number in each full block is stored
as a $b$-gap (block gap) rather than as a $d$-gap, computed as the
difference between this block's first document identifier and the
first document number in the previous block in the chain.
The $b$-gaps allow an indexed sequential access mode that supports
$\var{seek\_GEQ}(d)$ operations that scan a list for a given document
$d$, touching only the $b$-gap and $\var{n\_ptr}$ during the scan.
That is, the $b$-gaps provide support for {\emph{skipping}}, first
proposed by {\citet{mz96-tois}}.

With the vocabulary stored as a part of the index $\Index$, the
process of determining the correct head block for any given term must
also be handled carefully.
We employ a hash array of 32-bit integers that stores block offsets,
and map the characters of each term $t$ via that array to the offset
in $\Index$ of $t$'s head block.
We assume a hash array twice the size of the collection vocabulary
(using an extensible hashing technique), which then allows use of a
simple linear advance collision resolution technique, and thus
provides $O(|t|+1)$-time search, where $|t|$ is the number of
characters in $t$.
That is, if $v$ is the vocabulary size of the collection, the hash
array is costed at $8v$ bytes in our memory consumption results.
All of the compression results given in Section~\ref{sec-experiments}
include {\emph{every}} component of the index structures required.

As noted in Figure~\ref{fig-layout}, we make use of $h=4$-byte block
counts, limiting $\Index$ to $2^{32}$ blocks.
With $B=64$ a typical block size, that means that our in-memory
indexes are capped at $64 \times 2^{32}$ bytes, which is $256$~GiB.
The net implication of that limit is discussed further at the end of
Section~\ref{sec-space}.

\subsection{Adding Documents}
\label{sec-addingdocs}

As new documents arrive they are parsed into terms, and repeated
occurrences within the document collected together via a
sort-counting process.
Each term $t$ is also mapped via the hash array and using the
vocabulary information, to obtain the offset in $\Index$ of $t$'s
head block.

If the document contains any terms $t$ that have not appeared in
previous documents, an empty head block is allocated for each, by
taking the next unassigned block in $\Index$, indicated by a single
global counter that we denote as $\var{nblocks}$.
At the same time, {\var{nblocks}} gets stored into the position in
the hash table corresponding to $t$.
New head blocks $H$ then get their vocabulary components
assigned, and
have their tail block byte offset
write counters $H.\var{nx}$ initialized to $4h+2+|t|=18+|t|$, the
first location in $H$ that will be used for postings bytes.

\begin{algorithm}[t]
\caption{Adding a posting $\langle d, f\rangle$ for some term $t$ for
which the head block $H\equiv\Index[\var{h\_ptr}]$ has been
identified by the hash-based mapping.
Variable $\var{nblocks}$ is a global count of the number of blocks in
$\Index$ that are in use; and each $\var{n\_ptr}$ value requires
$h=4$ bytes.
\label{alg-addposting}}
\begin{algorithmic}[2]
\State
	{\bf{function}} $\var{add\_posting}(\Index, \var{h\_ptr}, d, f)$:
\State let $H$ represent the block $\Index[\var{h\_ptr}]$
	  \Comment{head block}
\State let $T$ represent the block $\Index[H.\var{t\_ptr}]$
	  \Comment{current tail block}
\State set $\var{gap} \leftarrow d - H.\var{last\_d}$
	\Comment{compute $d$-gap}
	\label{stp-addposting-dgap}
\State set $\var{nbytes} \leftarrow \var{code\_len}(\var{gap}, f)$
	\Comment{do a test encoding}
	\label{stp-addposting-testfit}
\If{$H.\var{nx} + \var{nbytes} > B$}
	\Comment{and check for fit}
	\label{stp-addposting-needgrowth}
  \State // need to allocate a new tail block, and convert
  	previous tail block to ``full''
  \State set $\var{gap} \leftarrow d - T.\var{d\_num}$
	\label{stp-addposting-findbgap}
	\Comment{use a $b$-gap at start of block}
  \State now let $F$ represent the block $\Index[H.\var{t\_ptr}]$
  \State and let $T$ represent the block $\Index[\var{nblocks}]$
	\label{stp-addposting-bgap}
  \State write null bytes from $F[H.\var{nx}]$ to $F[B-1]$ inclusive
  	\label{stp-addposting-zeroes}
  \State set $T.\var{d\_num} \leftarrow d$
  	\Comment{note first-in-block docnum}
  \State set $H.\var{t\_ptr} \leftarrow
  		F.\var{n\_ptr} \leftarrow \var{nblocks}$
	\label{stp-addposting-allocateblock}
  	\Comment{set block pointers}
  \State set $H.\var{nx} \leftarrow h$
  	\Comment{set the {\vbyte} write pointer}
	\label{stp-addposting-nx}
  \State set $\var{nblocks} \leftarrow \var{nblocks}+1$
  \State set $\var{nbytes} \leftarrow \var{code\_len}(\var{gap}, f)$
  \label{stp-addposting-endif}
\EndIf
\State code $\var{gap}$ and $f$ as bytes into $T$,
	starting at $T[H.\var{nx}]$
	\label{stp-addposting-write}
\State set $H.\var{nx} \leftarrow
		H.\var{nx} + \var{nbytes}$
	\Comment{advance the byte pointer}
\State set $H.\var{last\_d} \leftarrow d$
	\Comment{note most recent docnum}
\State set $H.\var{ft} \leftarrow H.\var{ft}+1$
	\label{stp-addposting-setft}
\end{algorithmic}
 \end{algorithm}

The list of unique terms associated with the new document and their
corresponding head block offsets is then processed via a sequence of
{\var{add\_posting}} operations, one per term, each of which appends
one posting to the index.
The process for inserting one new posting is shown in
Algorithm~\ref{alg-addposting}, with $B$ the uniform length of each
block in index $\Index$.
In the pseudo-code, variable $H\equiv\Index[\var{h\_ptr}]$
corresponds to the head block for term $t$, identifying the term this
posting is to be associated with; variable $T$ is the current tail
block and then the new tail block, should growth be required; and
variable $F$ denotes the previous tail block in those cases in which
growth gives rise to the need for a new tail block.
Either or both of $T$ and $F$ might indicate the same block as does
pointer~$H$.

The pseudo-code in Algorithm~\ref{alg-addposting} treats all of $H$,
$F$, and $T$ as being both compound structures (or rather, as
pointers to structures) and, when required, as plain arrays of $B$
bytes.
In particular, the ``{\verb+.+}'' operator is used to select the
fixed-width elements at the front of an indicated block (see
Figure~\ref{fig-layout}), with byte-by-byte access operations past
those fields carried out using the subscripting operator
``{\verb+[]+}'', counting bytes from the beginning of the block.
For example, with the first four fields in the head block $H$ each
stored as $h=4$-byte integers (see Figure~\ref{fig-layout}), the one-byte
element $H.\var{nx}$ could also be referred to as $H[16]$.
Note also that each block contains an integral number of postings,
and that our implementation does not split {\vbyte} values across
blocks -- the integrity of the $b$-gap support for indexed sequential
access was felt to be more desirable than this last small saving.
Unused tail bytes that get created by this decision are set to null,
allowing the {\vbyte} decoder to know that it has reached the end of
the block, as discussed in Section~\ref{sec-compression}.

Working through Algorithm~\ref{alg-addposting} in
detail, there are three phases to note.
In the first phase, lines~\ref{stp-addposting-dgap}
and~\ref{stp-addposting-testfit} compute the $d$-gap associated with
the new posting, and then calculate the length of its compressed
representation.
The second section of code, lines~\ref{stp-addposting-needgrowth}
to~\ref{stp-addposting-endif}, tests whether that posting can fit in
what remains of the current tail block
(step~\ref{stp-addposting-needgrowth}), and if it cannot, closes off
the current tail block by writing null bytes in all unused positions
(step~\ref{stp-addposting-zeroes}); allocates a new block and sets
the appropriate pointers (step~\ref{stp-addposting-allocateblock})
including the within-block byte counter
(step~\ref{stp-addposting-nx}); computes the required $b$-gap
(step~\ref{stp-addposting-findbgap}); and finally recalculates the
compressed posting size (step~\ref{stp-addposting-endif}).
The third phase, at steps~\ref{stp-addposting-write}
to~\ref{stp-addposting-setft} then writes the posting (as a $b$-gap
rather than a $d$-gap, if this is the first posting in the block)
into the tail block, and finally adjusts three variables in the head
block, thereby completing the {\var{add\_posting}} operation and
returning to the stable configuration shown in
Figure~\ref{fig-layout}.

\subsection{Double VByte -- Better Postings Compression}
\label{sec-magic}

\begin{algorithm}[t]
\caption{{\dvbyte}, packing the gap $g$ and frequency $f$ for a
posting into a single byte when $g<128/F$ and $f<F$, for threshold
parameter $F$.
\label{alg-magic}}
\begin{algorithmic}[2]
\State {\bf{function}} $\var{double\_vbyte\_encode}(g, f)$:
\If{$f<F$}
	\Comment{when the value $f$ is sufficiently small,}
  \State set $g' \leftarrow (g-1)\times F + f$
  	\Comment{$g$ and $f$ can be combined into a
			single value $g'$,}
  \State $\var{vbyte\_encode}(g')$
  	\Comment{and jointly coded using as few bytes
			as necessary}
\Else
  \State set $g' \leftarrow g\times F$
  	\Comment{or, when $f$ is larger than the threshold,}
  \State $\var{vbyte\_encode}(g')$
  	\Comment{an inflated version $g'$ of $g$
			is first coded,}
  \State $\var{vbyte\_encode}(f-F+1)$
  	\Comment{followed by a slightly adjusted version of $f$}
\EndIf
\State {\bf{return}}
\State
\State {\bf{function}} $\var{double\_vbyte\_decode}()$:
\State set $g' \leftarrow \var{vbyte\_decode}()$
  	\Comment{fetch a single value $g'$ from the stream,}
\If{$g' {\mbox{~{\bf{mod}}~}} F > 0$}
  	\Comment{and if it is not a multiple of $F$,}
  \State set $g \leftarrow 1 + g' {\mbox{~{\bf{div}}~}} F$ 
	\Comment{it carries both $g$ and $f$ embedded within it}
  \State set $f \leftarrow g' {\mbox{~{\bf{mod}}~}} F$
\Else
  \State set $g \leftarrow g' {\mbox{~{\bf{div}}~}} F$
	\Comment{or, when $g'$ is a multiple of $F$,
			it contains $g$ only}
  \State set $f \leftarrow F + \var{vbyte\_decode}()-1$
	\Comment{and $f$ was coded as a separate element}
\EndIf
\State {\bf{return}} $(g,f)$
\end{algorithmic}
 \end{algorithm}

Algorithm~\ref{alg-magic} describes another important part of our
proposal, a code-packing technique that can be used in any situation
in which byte codes are used to store postings.
The key idea is that when the $\fdt$ value is small (defined in the
pseudo-code as being less than the fixed value $F$, with $F=4$ a
typical value), instead of coding it independently and hence
consuming at least one byte to store each of the $\ddt$ and $\fdt$
values, the two components are folded together into a single value
before being coded.
The folding operation can be unambiguously reversed at decode time;
and is structured so that calls of the form $\var{vbyte\_encode}(0)$
can never arise, preserving the decoder's ability to know when to
stop decoding, as noted in Section~\ref{sec-compression}.
The comments embedded in the right-hand side of
Algorithm~\ref{alg-magic} provide further step-by-step guidance that
explains the details of the {\dvbyte} encoding and decoding
processes.
For example, when $F=4$, $g=10$ and $f=3$, a single byte covers the
packed value $g'=(g-1)\times F+f=39$, and the posting is one byte
shorter than had two separate {\vbyte} codes been
used, one for $g$ and one for $f$.
Then, when decoding, the value $g'=39$ is not a
multiple of $F=4$, and so the two components $g=10$ and $f=3$ can be
extracted out of that same single byte.

The {\dvbyte} code relies on there being a majority of cases in which
the computed value $g'$ is small enough that a one byte {\vbyte} code
suffices.
The combined value might also become large enough that a {\vbyte}
code of two bytes is needed, which is still no loss; this is what
happens if $g=40$ and $f=3$, with $g'=(g-1)\times F+f=159$ needing
two output bytes.
The pigeonhole principle means that there must also be cases in which
the $\ddt$ and $\fdt$ values would require one byte individually, but
in conjunction end up requiring three bytes.
An example of this third situation occurs when $g=40$ and $f=5$, with
$g'=160$ needing to be coded (two bytes) followed by $f-F+1=2$ in a
third byte.

\subsection{Preliminary Results}
\label{sec-prelimresults}

Fortunately, the typical frequency distributions associated with
postings data -- a very high fraction of low values of $f$; many
small values of $g$; and a joint distribution that has larger values
of $g$ highly likely to be accompanied by low values of $f$ -- mean
that the third of those three cases is relatively rare and that it is
the first case that dominates.
Table~\ref{tbl-byte-byebye-dgaps} provides evidence in support of
that claim, derived from the {\wsj} collection of approximately
$100{,}000$ documents and $21$ million postings (see
Table~\ref{tbl-datasets} in Section~\ref{sec-methodology} for details
of the three test collections used in our experiments).
It shows the breakdown of postings costs, categorized by initial size
across the columns as the sum of two separate calls to
$\var{encode\_vbyte}()$, and then by transformed size via a single
call $\var{encode\_double\_vbyte}(g,f)$ within each of the
columns.
The percentages in each column below the line add up to the number in
the ``before'' row above the line, with all percentages relative to
the total number of postings.
The overwhelming dominance of the blue values (one byte saved)
compared to the corresponding red values (one byte extra required)
demonstrates the usefulness of the new technique.
In this collection there were no five-byte postings required using
either mechanism.

\begin{table}[t]
\centering
\renewcommand{\tabcolsep}{1.5em}
\newcommand{\tabent}[1]{\makebox[8mm][c]{#1}}
\sisetup{
group-separator = {,},
round-mode = places,
round-precision = 2,
table-format=2.2,
}\begin{tabular}{c c SSSS}
\toprule
\multirow{1}{*}{New size}
	&& \multicolumn{4}{c}{Original posting size ({\vbyte}, bytes)}
\\
\cmidrule{3-6}
{({\dvbyte},}
	&& {\tabent{2}}
		& {\tabent{3}}
			& {\tabent{4}}
				& {\tabent{5}}
\\
{bytes)}
& & 82.192425\% & 16.822099\% & 0.985476\% & {--}
\\
\midrule
1 & & \color{blue} 58.505458\% & {--} & {--} & {--}
\\
2 & & 22.922610\% & \color{blue} 14.565331\% & {--} & {--}
\\
3 & & \color{red} 0.764357\% & 2.181501\% & \color{blue} 0.933330\% & {--}
\\
4 & & {--} & \color{red} 0.075267\% & 0.052146\% & {--}
\\

\bottomrule
\end{tabular}\aftertabspace
 \caption{Percentages of $\dtdf$ postings of each given length in
the {\wsj} collection when represented as separate {\vbyte} codes for
the $d$-gap and the $\fdt$ component, and the distribution of
{\dvbyte} sizes when jointly coded via Algorithm~\ref{alg-magic}
using $F=4$.
Table~\ref{tbl-datasets} provides details of the document collection.
\label{tbl-byte-byebye-dgaps}}
\end{table}

\begin{table}[t]
\centering
\renewcommand{\tabcolsep}{0.8em}
\sisetup{
group-separator = {,},
round-mode = places,
round-precision = 3,
table-format = 1.3,
}\begin{tabular}{l SSSSS
}
\toprule
& {$F=1$}
	& {$F=2$}
		& {$F=4$}
			& {$F=8$}
				& {$F=16$}
\\
\midrule
{Bytes/posting}
	& 2.187931
	& 1.578910
	& \color{blue} 1.456286
	& 1.501356
	& 1.603331
\\
{Ratio}
	& 1.000000
	& .721645
	& \color{blue} .665600
	& .686199
	& .732807
\\
\bottomrule
\end{tabular}\aftertabspace
 \caption{{\dvbyte} postings costs measured in bytes
per posting and relative size ratio for the {\wsj} collection, as a
function of transformation parameter $F$.
When $F=1$ the original {\vbyte} scheme results.
\label{tbl-savings}}
\end{table}

Table~\ref{tbl-savings} translates those behavioral patterns into
concrete costs, expressed now in units of bytes per posting (and in
this preliminary table, counting postings only, and not including the
hash array, nor any of the fixed costs associated with the index
blocks).
When $F=1$ the original ``separate {\vbyte} codes'' scheme is in
operation; as can be seen, relative to that baseline fully one third
of the postings cost can be eliminated by {\dvbyte} when $F=4$.
We use $F=4$ in all subsequent experiments on document-level inverted
indexes.

We also carried out a preliminary
``straight through'' speed test experiment, to compare {\vbyte} and
{\dvbyte} encoding and decoding rates.
Table~\ref{tbl-straightspeed} summarizes the outcomes.
In this experiment, the complete set of postings of the {\wiki} test
collection (see Section~\ref{sec-methodology} for details) was placed
into an array of $32$-bit integers in memory, as a sequence of
alternating $d$-gaps and $\fdt$ values, without any further metadata
of any sort.
They were then encoded to a second array of unsigned bytes, measuring
the time taken.
That second array was then decoded back to a third array of $32$-bit
integers also in memory; and finally, the first and third arrays
compared, to verify the integrity of the process.
A parameter of $F=4$ was used in {\dvbyte}.

\begin{table}[t]
\centering
\renewcommand{\tabcolsep}{1.0em}
\sisetup{
group-separator = {,},
round-mode = places,
round-precision = 3,
table-format = 1.3
}\begin{tabular}{
l
	S
		S
			S
}
\toprule
Method
	& {Encoding}
		& {Decoding}
			& {Effectiveness}
\\
\midrule
{\vbyte}
	& 2.84432
		& 2.98146
& 2.303839
\\
{\dvbyte}
	& 3.45124
		& 3.55526
& 1.628734
\\
Copy, looping over bytes
	& 1.68028
		& 2.14913
			& {8}
\\
Copy, looping over integers
	& 1.28229
		& 1.13503
			& {8}
\\
{\tt{memcpy}}, whole slab at once
  & 1.347497
    & 1.14495
      & {8}
\\
\bottomrule
\end{tabular}\aftertabspace

\caption{Array-based encoding and decoding speed,
measured over the $\num[round-precision=0]{996277511}$ document-level
postings of the form~$\dtdf$ extracted from the {\wiki} test
collection; that is, approximately $\num[round-precision=2]{1.99255}$
billion integers.
The {\dvbyte} parameter $F$ was set to four; en/decoding times are
measured in seconds for the complete set of postings (smaller is
better); and compression effectiveness is measured in compressed
bytes per posting (smaller is better).
The experimental hardware is described in
Section~\ref{sec-methodology}.
\label{tbl-straightspeed}}
\end{table}

As can be seen from the table, the {\dvbyte} mechanism
is approximately $20$\% slower than is {\vbyte}, for both encoding
and decoding; with the differential caused by the need for additional
control logic, including an if-statement and possible branch (see
Algorithm~\ref{alg-magic}).
In turn, both are slower than plain copy operations, in which bytes
or whole words are transferred via a loop.
In the last line of the table the standard function {\tt{memcpy()}}
is used, which is about three times faster than are {\vbyte} and
{\dvbyte}.
The right-most column of Table~\ref{tbl-straightspeed} provides the
justification for the use of {\vbyte} and, even better, {\dvbyte},
confirming that relative to plain integers, {\dvbyte} saves almost
$80$\% of the postings cost.

\subsection{Querying Operations}

The index structure shown in Figure~\ref{fig-layout} is always
available for querying, and search operations can be interleaved with
document insertions.
To resolve a query $Q$ the hash array and term strings stored in the
corresponding entries in $\Index$ are used to locate the set of
active head blocks, one per query term, indicated by a set of
$\var{h\_ptr}$ variables.
Each of those head blocks contains the first tranche of postings for
that term, with more blocks linked via the $\var{n\_ptr}$ fields.
Within any given block posting decoding ends if a null byte is
detected, or if the $B$\,th byte has been reached, at which time the
next block is accessed via that current block's stored
$\var{n\_ptr}$.
Decoding of the entire posting chain for any term ends when the block
$\Index[\var{h\_ptr}].\var{t\_ptr}$ is reached, and then, within that
tail block, when the first $\Index[\var{h\_ptr}].\var{nx}$ bytes of
it have been decoded.

Boolean search modes, and ranking models that use $\fdt$ values only,
need no additional support beyond that which has been captured in
Figure~\ref{fig-layout}, and we explore two such options in
Section~\ref{sec-experiments}.
More complex querying can also be supported, provided that any
additional necessary information is also made available.
For example, ranked querying models in which a document-length
normalization step is required must maintain a separate array of
document lengths; we consider that to be not part of the core
inverted index, and do not include it in our space measurements.
Similarly, ranking modes that employ dynamic pruning heuristics need
a per-term upper bound; that would require allocation of an
additional $4$-byte field in each head block, slightly increasing the
size of the resultant index.
 \section{Experiments}
\label{sec-experiments}

We now describe a range of more detailed experiments that capture the
balance of performance (see Figure~\ref{fig-sausages}) that is
possible using our index structure.

\subsection{Data and Hardware}
\label{sec-methodology}

Our experiments make use of three widely-accessible text collections:
\begin{itemize}
  \item {\wsj} is the first half of the {\emph{Wall Street Journal}}
    collection, consisting of
    newspaper text from 1987--1989, as used in several of the 
    very early TREC activities.

  \item {\robust} is a collection of newswire documents from the 
    1980s and 1990s. It is made up of TREC Disks 4 and 5, excluding the 
    {\emph{Congressional Record}} from Disk 4.

  \item {\wiki} is a dump of the English Wikipedia corpus from April
  2, 2022.
  Documents were extracted using the
  {\method{WikiExtractor}}\footnote{\url{https://github.com/attardi/wikiextractor}}
  tool.

\end{itemize}

Each collection was first converted to a common {\emph{docstream}}
format prior indexing.
A docstream represents documents as single lines of text, with the
first element a document identifier, and the remainder of the line an
ordered set of terms making up that document.
A series of simple pre-processing steps was also applied as the
docstreams were created: long terms were broken after each group of
$20$ consecutive alphabetic characters; sequences of non-alphabetic
characters were replaced with single spaces; and uppercase characters
were folded to lowercase.
No particular attention was applied to SGML or HTML markup elements,
and the alphabetic components of those tags were retained in the
docstreams.
No stemming or stopping was performed either.
Table~\ref{tbl-datasets} provides statistics of the resultant
docstream collections that were then used in our experimentation.
The largest of the three collections, {\wiki}, included more than two
billion words, and gave rise to document-level indexes containing
nearly a billion postings.

\begin{table}[t]
\centering
\renewcommand{\tabcolsep}{0.4em}
\begin{tabular}{
l
	S[round-precision=0,table-format=10.0]
		S[round-precision=0,table-format=9.0]
			S[round-precision=0,table-format=7.0]
				S[round-precision=2,table-format=1.2]
					S[round-precision=1,table-format=3.1]
  					S[round-precision=1,table-format=5.1]
}
\toprule
	& {Words}
		& {Postings}
			& {Documents}
				& {Words/post.}
					& {Words/doc.}
           & {Size (MiB)}
\\
\midrule
{\wsj}
	& 42899195
		& 20716886
			& 98732
				& 2.070735
					& 434.501428
            & 239.925967
\\
{\robust}
  & 278511531
    & 121987739
     & 528155
       & 2.2831108
         & 527.329157
          & 1551.2772865
\\
{\wiki}
& 2444347604 & 996277511
			& 6477362
				& 2.453480
					& 377.367762
           & 13839.08490848
\\
\bottomrule
\end{tabular}\aftertabspace
 \caption{Datasets used in experiments, after pre-processing.
The final column refers to the docstreams prepared from the input
text.
\label{tbl-datasets}}
\end{table}

\begin{table}[t]
\centering
\renewcommand{\tabcolsep}{0.7em}
\begin{tabular}{
l
	S[round-precision=0,table-format=5.0]
		S[round-precision=3,table-format=2.3]
			S[round-precision=0,table-format=5.0]
				S[round-precision=0,table-format=7.0]
}
\toprule
	& {Queries}
    & {Avg.\ Length}
			& {Avg.\ Conj.\ Match}
				& {Avg.\ Disj. Match}
\\
\midrule
{\wsj}
	& 28104
		& 2.87923
			& 339.87
				& 28336.2
\\
{\robust}
  & 28104
    & 2.87923
     & 2001.29
       &  155554.0
\\
{\wiki}
	& 28104 
		& 2.87923
			& 16544.5
				& 1502997.232
\\
\bottomrule
\end{tabular}\aftertabspace
 \caption{Queries used in experiments, after the filtering step.
Every query that was retained had at least one conjunctive match in
each of the three collections.
\label{tbl-queries}}
\end{table}

To generate test queries, all $60{,}000$ queries from the TREC
Million Query Track {\cite{ac+07-trec, aa+08-trec, cp+09-trec}} were
used as a starting point.
Any queries that did not have a conjunctive match in each of the
three test collections was then removed, to obtain a filtered log of
around $28$ thousand queries; the properties are summarized in
Table~\ref{tbl-queries}.
The last column gives the average number of documents that contained
{\emph{any}} of the terms in each query; the second-to-last the
average number of documents that contained {\emph{every}} query term.

Our experimental software was implemented in C++ and compiled with
{\tt{gcc 7.5.0}} with {\tt{-O3}} optimization.
All experiments were conducted on a Linux server with two $3.50$ GHz
Intel Xeon Gold 6144 CPUs and $512$ GiB of RAM.
Indexing experiments used an $894$ GiB SATA SSD with transfer speeds
of up to $5.6$ GiB/s for all I/O.
All experiments used a single processing core.

\subsection{Experimental Objectives}

We are now in a position to investigate the research
goals that were listed in Section~\ref{sec-mode}.
In particular, we restrict our attention to the dynamic
immediate-indexing part of a larger retrieval system that might also
have some static index components (see Figure~\ref{fig-mode}), and
focus on an incoming stream of documents, so as to measure the
indexing process and then the cost of querying against that dynamic
index.
Given that intention, the high-effectiveness compression regimes
described in Section~\ref{sec-compression} cannot be employed, and
nor can the query execution mechanisms described in
Section~\ref{sec-pruning}.
Both are relevant to the static shard index components, but not to
the dynamic part.
Focusing on the dynamic index component also means that we need to
consider a suitable scale, and that is the purpose of the three
selected datasets, each of which constitutes a plausible batch of
updates in the context of a much larger retrieval system.
For example, the {\wiki} collection contains around a billion
postings, and an uncompressed index for it would require
approximately $8$~GiB of memory; it can be thought of as being
$10$\%, or $1$\%, or even $0.1$\% of a ``whole'' retrieval system
with the structure shown in Figure~\ref{fig-mode}.
Given that organization, and the fact that all aspects of our new
mechanism scale linearly, the three test collections support
experiments that are realistic in terms of the proposed operational
environment.

Note also that while we primarily
explore document retrieval using the standard document-at-a-time
processing strategy, there is nothing in the new mechanism that
precludes the use of term-at-a-time processing, since both options
share the same document-sorted index requirement.
On the other hand, score-at-a-time index processing is based on a
different index organization, and would not be an option.
Finally, note that the query time measurements reported below form an
important complement to the ``straight through'' speeds already
documented in Section~\ref{sec-prelimresults}, and help to locate the
new mechanism in the trade-off space illustrated in
Figure~\ref{fig-sausages}.

\subsection{Baselines}

Two previous implementations are used as reference systems.
As a competitive baseline for both index size and query processing
latency when a static index can be employed, and thus covering two of
the corners of Figure~\ref{fig-sausages}, we use the C++
{\method{PISA}} system {\cite{pisa19-osirrc}}.
Two specific configurations were employed: one aimed at minimizing
index space consumption via block-based interpolative coding
{\cite{ms00-irj}} (denoted {\method{PISA-Interp}}); and one that
provides a good balance between access time and space occupancy via
the SIMD-BP128 bitpacking codec {\cite{lb15-spe}} (denoted
{\method{PISA-BP128}}).
The {\method{PISA}} software has been shown to be one of the fastest
query processing systems available
{\cite{mm20cikm,pisa19-osirrc,mss19-ecir}}, and has also been
independently compared to other open-source systems in terms of
conjunctive Boolean query speed.\footnote{
{\url{https://tantivy-search.github.io/bench/}}, accessed 30 May
2022.}
It thus provides an aspirational reference point for the new dynamic
system in terms of both index size and query speed.

Then, in separate experiments that are presented in
Section~\ref{sec-likeforlike}, we compare the new immediate-access
indexing approach to the previous in-memory mechanism of
{\citet{hb17adcs}}.
This second round of experimentation complements the aspirational
{\method{PISA}} ones presented shortly, and allows a direct comparison
with an alternative system which is also designed to operate in the
upper half of the space shown in Figure~\ref{fig-sausages}.

\subsection{Index Size}
\label{sec-space}

\begin{table}[t]
\centering
\begin{tabular}{l@{~}l
	cS[round-precision=0,table-format=6.0]
		S[round-precision=0,table-format=8.0]
			S[round-precision=1,table-format=3.1]
	cS[round-precision=0,table-format=6.0]
		S[round-precision=0,table-format=8.0]
			S[round-precision=1,table-format=3.1]
}
\toprule
\multicolumn{2}{l}{\multirow{2}{*}{Block and element}}
	&& \multicolumn{3}{c}{$B=48$ bytes}
		&& \multicolumn{3}{c}{$B=64$ bytes}
\\
\cmidrule{4-6}\cmidrule{8-10}
	&&& {Blocks}
		& {Bytes}
			& {Percent}
	&& {Blocks}
		& {Bytes}
			& {Percent}
\\
\midrule
\multicolumn{2}{l}{Head blocks}
	&& 159734
	&&&& 159734
\\
& -- link pointers
	&&
		& 1277872
			& 3.0805\%
	&&
		& 1277872
			& 3.0056\%
\\
& -- vocabulary
	&& 
		& 2817304
			& 6.7916\%
	&& 
		& 2817304
			& 6.6264\%
\\
& -- postings
	&&
		& 1711686
			& 4.1263\%
	&&
		& 2362446
			& 5.5566\%
\\
& -- trailing null bytes
	&&
		& 1860370
			& 4.4847\%
	&&
		& 3765354
			& 8.8563\%
\\
\multicolumn{2}{l}{Full blocks}
	&& 632105
	&&&& 449644
\\
& -- link pointers
	&&
		& 2528420
			& 6.0952\%
	&&
		& 1798576
			& 4.2303\%
\\
& -- postings
	&&
		& 27628440
			& 66.6028\%
	&&
		& 26851357
			& 63.1555\%
\\
& -- trailing null bytes
	&&
		& 184180
			& 0.4440\%
	&&
		& 127283
			& 0.2994\%
\\
\multicolumn{2}{l}{Tail blocks}
	&& 45755
	&&&& 34972
\\
& -- docnums
	&&
		& 183020
			& 0.4412\%
	&&
		& 139888
			& 0.3290\%
\\
& -- postings
	&&
		& 925397
			& 2.2308\%
	&&
		& 955899
			& 2.2483\%
\\
& -- unused bytes
	&&
		& 1087823
			& 2.6224\%
	&&
		& 1142421
			& 2.6608\%
\\
\multicolumn{2}{l}{Hash array}
	&&
		& 1277872
			& 3.0805\%
	&&
		& 1277872
			& 3.0056\%
\\
\multicolumn{2}{l}{Total size}
	&& 837594
		& 41482384
			& 100.0\%
	&& 644350
		& 42516272
			& 100.0\%
\\
\bottomrule
\end{tabular}\aftertabspace
 \caption{Blocked index components (see Figure~\ref{fig-layout}) and
their relative contributions when indexing the {\wsj} collection,
using two different block sizes $B$.
All percentages are relative to the corresponding total index size in
the last row, which includes the hash array, all vocabulary and
global term statistics, all of the postings, and all overheads
associated with the extensible lists including unused bytes at the
end of allocated blocks.
If a postings list contains only one block that block is counted as a
head block, with the list not having a tail block.
\label{tbl-components}}
\end{table}

Table~\ref{tbl-components} gives an insight to the cost of the
various components that make up a document-level index for {\wsj},
for two typical values of the block size $B$.
The dominant cost is the postings component of the full blocks, and
they are where the heavy lifting gets done.
Our design decision to not split postings across blocks has resulted
in a small overhead cost in terms of unused bytes at the end of those
blocks (around $0.3$\%--$0.4$\% of total space) but also made both
indexing and querying substantially more straightforward.
Note also that more than two thirds of the terms never ``escape''
their head block to create a first tail block, and that the unused
space at the end of head blocks is a non-trivial overhead, especially
when $B=64$.
Indeed, there is more unused space in the head blocks than in the
tail blocks, even though the latter are on average only half full,
and even though the former also contain the term's vocabulary
information.

\begin{table}[t]
\centering
\renewcommand{\tabcolsep}{0.6em}
\sisetup{
table-format=1.3,
round-precision=3,
}
\begin{tabular}{ll @{\hspace{2em}}  SSSSSS}
\toprule
\multicolumn{2}{l}{Collection}
	& {$B=40$}
		& {$B=48$}
			& {$B=56$}
				& {$B=64$}
					& {$B=72$}
						& {$B=80$}
\\
\midrule
{\wsj} & doc-level
	& \color{blue} 1.99977
		& 2.00235
			& 2.0220
				& 2.05225
					& 2.08752
						& 2.12739
\\
{\robust} & doc-level
	& 1.86334
    & \color{blue} 1.85921
			& 1.8701
				& 1.88919
					& 1.91356
						& 1.9411
\\
{\wiki} & doc-level
	& 2.10733
    & \color{blue} 2.09138
			& 2.09557
				& 2.10995
					& 2.13057
						& 2.15542
\\
\bottomrule
\end{tabular}\aftertabspace
 \caption{Compression effectiveness for document-level indexes as
block size $B$ varies, in bytes per posting in all cases.
Encoding parameter $F=4$ was used throughout.
All sizes include the hash table, terms and all other vocabulary
information, plus the postings themselves.
Block sizes less than $40$ cannot be used.
The best value in each row is shown in blue.
\label{tbl-blocksize}}
\end{table}

Table~\ref{tbl-blocksize} provides compression effectiveness results
for all three collections and a range of typical values of $B$.
These results are expressed as bytes per posting, and if required can
be converted to megabyte sizes using the information provided in
Table~\ref{tbl-datasets}.
The listed compression rates include all aspects of the index
(accounted for in detail in Table~\ref{tbl-components} for two of the
{\wsj} values in the first row), and provide an accurate assessment
of the memory footprint that would be required if dynamic indexes had
been constructed from the corresponding docstreams.
The blue values show the best compression rates for each collection,
with the variation away from those minima being relatively modest as
$B$ is varied.
The {\dvbyte} compression technique (Algorithm~\ref{alg-magic}) is an
important contributor to the compact indexes that are created, along
with the careful use (and re-use) of memory as detailed in
Figure~\ref{fig-layout} and Algorithm~\ref{alg-addposting}.

\begin{table}[t]
\centering
\begin{tabular}{
ll
	S[round-precision=3,table-format=1.3]
		S[round-precision=3,table-format=1.3]
}
\toprule
Collection &
	& {\method{PISA-Interp}}
		& {\method{PISA-BP128}}
\\
\midrule
{\wsj} & doc-level 
	& 1.1619726
		& 1.5137522
\\
{\robust} & doc-level
	& 1.0752407
		& 1.5151313
\\
{\wiki} & doc-level
	& 1.3183456
		& 1.7996905
\\
\bottomrule
\end{tabular}\aftertabspace
 \caption{Compression effectiveness for document-level
{\method{PISA}} indexes in bytes per posting, using two different
compression regimes for postings, and including vocabulary and other
files required for conjunctive querying.
These results for static indexes can be compared against the
corresponding dynamic index costs shown in Table~\ref{tbl-blocksize}.
\label{tbl-pisa-space}}
\end{table}

Table~\ref{tbl-pisa-space} provides a reference point against which
the results of Table~\ref{tbl-blocksize} can be assessed.
These come from the {\method{PISA}} system using the two different
compression plug-ins {\cite{mss19-ecir}} described above, and include all 
of the components needed to run a static in-memory retrieval system.
As anticipated, both modalities attain better compression than is
shown in Table~\ref{tbl-blocksize}, showing the benefit of knowing
each posting list's characteristics and being able to tune
compression parameters to match.
The static {\method{PISA}} indexes also have no unused end-of-block
space, nor any link pointers, both of which are inevitable in a
dynamic index (see Table~\ref{tbl-components}).
Indeed, the relatively small ratio between the results in
Table~\ref{tbl-pisa-space} and Table~\ref{tbl-blocksize} is a
positive outcome that indicates that the memory overhead incurred by
dynamic indexing need not be excessive.

Combining the information in Tables~\ref{tbl-datasets}
and~\ref{tbl-blocksize} indicates that a dynamic index of $256$~GiB
(the upper limit possible with $B=64$ and $h=4$) will be able to
support more than $100$ billion postings, corresponding to
approximately $200$ billion words in typical-length documents, and
hence to more than $1$~TiB of plain text.

\subsection{Indexing Throughput}

\begin{figure}[t]
\centering
\includegraphics[width=0.7\textwidth]{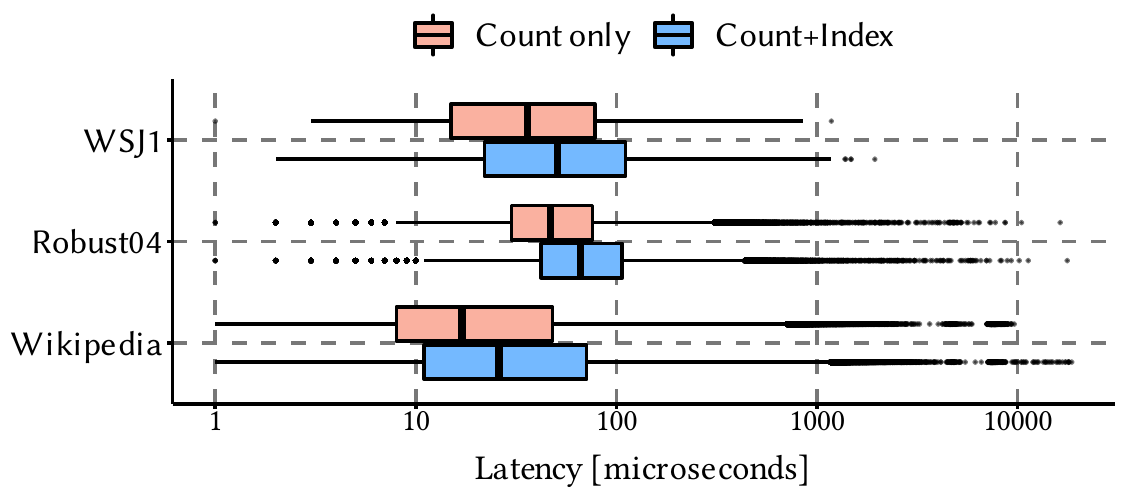}
\caption{Per-document insertion time in microseconds, with
``{\method{Count only}}'' the time taken to tokenize and count the
terms in each document without inserting them into the index, as a
lower-bound on indexing cost; and with ``{\method{Count+Index}}''
then including the cost of inserting the postings as well.
\label{fig-insert}}
\end{figure}

Figure~\ref{fig-insert} shows how fast indexing is with our proposed
mechanism, reporting indexing times
measured in units of microseconds per document.
Each collection was processed twice: a first time with all of the
calls to $\var{add\_posting}()$ (see Algorithm~\ref{alg-addposting})
returning immediately, without doing any work, shown as
``{\method{Count only}}''; and then a second time with the
$\var{add\_posting}()$ calls all fully functional.
The difference in the two times is then cost of index construction,
with the balance of the time being required to read the input, parse
it into tokens, and count the term frequencies within each document
-- processing that is required by all indexing implementations.

The $\var{add\_posting}()$ calls take only a small fraction of the
indexing time.
For example, on {\wiki} the {\method{Count only}} time is $420.7$
seconds, and {\method{Count+Index}} was $574.5$ seconds, or over a
gigabyte per minute.
As further reference points, {\citet{envelope}} reported the
production-ready Lucene system to achieve up to $5.1$ GiB per minute
when indexing the {\method{ClueWeb12B}} collection using $48$
processing threads, including tokenization; and {\citet{hb17adcs}}
report a ``List Building'' time of around $320$ seconds for a
collection of approximately one billion word-level postings (their
Table~7, row ``FibonacciB, Large Pages'').
The latter value is validated in Section~\ref{sec-likeforlike}.

\subsection{Query Throughput}
\label{sec-queryspeed}

We investigate response latency using two separate query modes:
Boolean conjunctions in a document-at-a-time manner (see
{\citet[Algorithms~1 and~3]{cm10-tois}}), returning a list of all
matching document identifiers; and via a document-at-a-time top-$k$
disjunctive retrieval strategy.
In the latter mode the top-$k$ documents ``seen so far'' are tracked
in a min-heap structure, with each document that contains any of the
query terms scored and compared against the smallest element in the
heap.
In the {\method{PISA}} baseline system scores were computed according
to the common {\method{BM25}} model {\cite{rz09fntir}} (the precise
formulation used can be found in the {\method{PISA}} overview
{\cite{pisa19-osirrc}}).
In our system we employed a simple TF$\times$IDF ranking model in
which the weight $w$ of term $t$ in document $d$ is computed as
\[
	w_{t,d} = \log(1 + f_{t,d}) \times \log(1 + N/f_t) \, ,
\]
where $f_{t,d}$ is the frequency of $t$ in $d$, where $f_t$ is the
number of documents that contain $t$, and where $N$ is the number of
documents in the collection; and created answer sets containing the
top $k=10$ answers for each query.
Note that our interest here is on retrieval speed, and that we are
not claiming that the similarity formulation employed is competitive
in terms of retrieval effectiveness.
Note that this experiment is not designed to compare like-with-like.
As already noted, the {\method{PISA}} comparator system makes use of
pre-computed and pre-optimized static indexes; in addition, there are
also differences in the two similarity formulations that might affect
their relative speed.
In particular, when the {\method{BM25}} computation
embedded in {\method{PISA}} was replaced by a TF$\times$IDF version,
query processing became approximately $25$\% slower.
When that
relativity is taken into account, this experiment gives helpful
guidance in regard to the difference in cost between querying using
an optimized static index, and querying using our dynamic index,
which is designed for versatility.
In a full system both types of querying might be required, as is
shown in Figure~\ref{fig-mode}.

\begin{figure}[t]
\centering
\includegraphics[width=\textwidth]{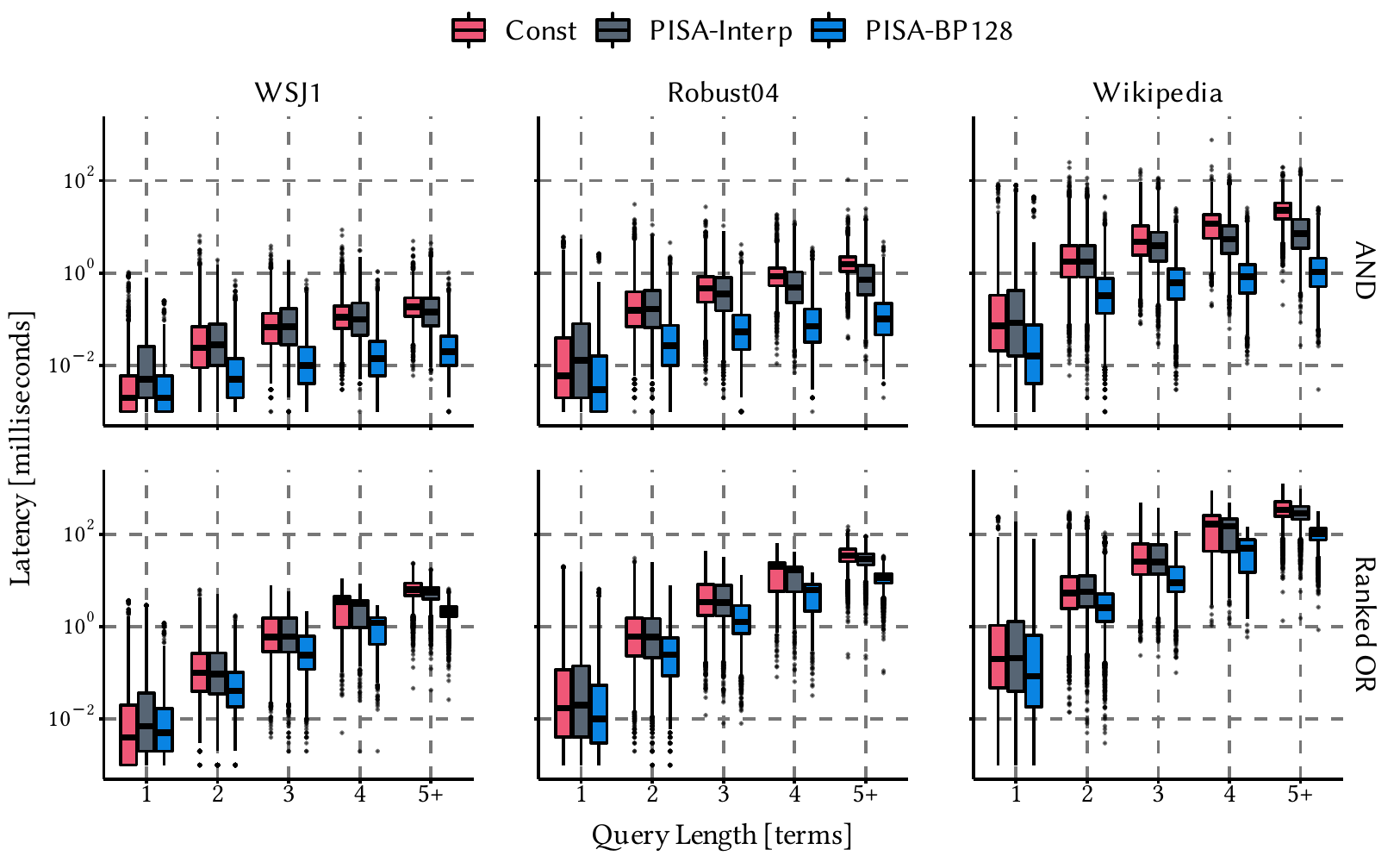}
\caption{Query processing time distribution in milliseconds per
query, using the filtered MQT query log (Table~\ref{tbl-queries}) and
plotted as a function of the number of terms in each query.
Each graph compares our dynamic index with two {\method{PISA}}-based
systems, with the six graph panes covering three collections
(Table~\ref{tbl-datasets}) and two querying modes (conjunctive
Boolean, and top-10 disjunctive).
\label{fig-qp}}
\end{figure}

Figure~\ref{fig-qp} shows per-query query latency distributions for
two versions of the {\method{PISA}} reference system,
in both cases making use of static
indexes, static querying modes,
and the faster {\method{BM25}} computation,
comparing them to our proposed dynamic index
arrangement.
The six panes represent three collections and two querying modes.
While both {\method{PISA}}-based systems are (as expected) faster
than ours, nor is our system a laggard.
For example, conjunctive queries on the {\robust} collection can be
evaluated in around $550$ microseconds on average for queries of up
to four terms, and the more expensive ranked disjunctive queries of
the same length can be resolved in around five milliseconds on
average.
 \section{Extensions}
\label{sec-theresmore}

In this section we consider several extensions to the scheme
described in Section~\ref{sec-somethingnew} and measured in
Section~\ref{sec-experiments}.
In particular, we consider word-level indexing, showing that
{\dvbyte} is again a useful technique, albeit with a twist
(Section~\ref{sec-wordlevel}); we consider variable block sizes,
introducing a new extensible list approach that allows an asymptotic
reduction in the overhead cost ratio (Sections~\ref{sec-varblocks}
and~\ref{sec-triangle});
and we describe a simple block rearrangement mechanism that
accelerates query processing speed and involves only a minor
interruption to the document ingestion process
(Section~\ref{sec-rearranging}).

\subsection{Word-Level Indexing}
\label{sec-wordlevel}

The structure and approach described in
Section~\ref{sec-somethingnew} can be used to construct word-level
indexes with only small modifications required.
Postings $\dtdw$ (see Table~\ref{tbl-indexes}) in which $\ddt$ is the
ordinal number of the $i$\,th document containing term $t$, and
$\wdt$ is the ordinal word number within $d$ of the $j$\,th instance
of $t$, are coded as two values.
The first value is a $d$-gap relative to the immediately preceding
posting, but then with one added, to ensure that the resultant $\gdt$
value is strictly greater than zero (Section~\ref{sec-compression}).
The second element in each posting is a corresponding $w$-gap that
represents the ordinal word number, if this is the first occurrence
of $t$ in $\ddt$, or represents the interval since the most recent
previous instance of $t$, if this is $\ddt$'s second or subsequent
occurrence of $t$.
In this arrangement the term occurrence count $\fdt$ becomes a
derived quantity, calculable by scanning the postings list.

In a word-level index every term in the input generates a posting.
That means that there are many $d$-gaps of just one, arising when the
term appears multiple times in the same document; and also a smaller
but still significant fraction for which the $d$-gap is two,
corresponding to $t$ appearing in consecutive documents.
On the other hand, the $w$-gaps tend to be larger, since it is
relatively rare for a word to closely follow itself in a document
(with the exception of some song lyrics, ``she loves you, yeah, yeah,
yeah'').

\begin{table}[t]
\centering
\renewcommand{\tabcolsep}{0.8em}
\newcommand{\tabent}[1]{\makebox[8mm][c]{#1}}
\sisetup{
group-separator = {,},
round-mode = places,
round-precision = 2,
table-format=2.2,
}\begin{tabular}{c c SSSS}
\toprule
\multirow{1}{*}{New size}
	&& \multicolumn{4}{c}{Original posting size ({\vbyte}, bytes)}
\\
\cmidrule{3-6}
{({\dvbyte},}
	&& {\tabent{2}}
		& {\tabent{3}}
			& {\tabent{4}}
				& {\tabent{5}}
\\
{bytes)}
& & 56.166802\% & 36.906623\% & 6.566310\% & 0.360266\%
\\
\midrule
1 & & \color{blue} 30.223775\% & {--} & {--} & {--}
\\
2 & & 18.966911\% & \color{blue} 14.300189\% & {--} & {--}
\\
3 & & \color{red} 6.976117\% & 21.107254\% & \color{blue} 0.052041\% & {--}
\\
4 & & {--} & \color{red} 1.499180\% & 6.430953\% & \color{blue} 0.000044\%
\\
5 & & {--} & {--} & \color{red} 0.083316\% & 0.360030\%
\\

\bottomrule
\end{tabular}\aftertabspace
 \caption{Percentages of $\dtdw$ postings of each given length in
the {\wsj} collection when represented as separate {\vbyte} codes for
the $\ddt$ and the $\wdt$ components, and the distribution of altered
sizes when jointly coded via Algorithm~\ref{alg-magic}, using $F=3$.
\label{tbl-byte-byebye-wgaps}}
\end{table}

The {\dvbyte} approach can again be used to create compound codes,
but should be applied with the arguments swapped (the ``twist''
mentioned above), so as to exploit that reversal of the two relative
frequency distributions.
In our word-level indexes, an adjusted $\langle g,w\rangle$
combination is encoded using $\var{double\_vbyte\_encode}(w,g)$ (and
using $F=3$), whereas in the document-level indexes considered in
Sections~\ref{sec-somethingnew} and~\ref{sec-experiments} each
$\langle g,f\rangle$ combination was represented via a call to
$\var{double\_vbyte\_encode}(g,f)$.
Table~\ref{tbl-byte-byebye-wgaps} demonstrates that this minor
adaptation again leads to substantial compression savings, with
around $45$\% of the postings becoming one byte shorter and less than
$9$\% becoming one byte longer.

\begin{table}[t]
\centering
\renewcommand{\tabcolsep}{0.6em}
\sisetup{
table-format=1.3,
round-precision=3,
}
\begin{tabular}{ll @{\hspace{2em}}  SSSSSS}
\toprule
\multicolumn{2}{l}{Collection}
	& {$B=40$}
		& {$B=48$}
			& {$B=56$}
				& {$B=64$}
					& {$B=72$}
						& {$B=80$}
\\
\midrule
{\wsj}
& word-level
	& 2.61361
		& 2.57096
			& 2.55022
				& \color{blue} 2.54215
					& \color{blue} 2.54225
						& 2.5480
\\
{\robust}
& word-level
	& 2.41398
		& 2.37044
			& 2.34643
				& 2.33379
          & \color{blue} 2.32813
            & \color{blue} 2.32751
\\
{\wiki}
& word-level
	& 2.49901 
		& 2.4484
			& 2.41999
				& 2.40379
					& 2.39534
            & \color{blue} 2.39214
\\
\bottomrule
\end{tabular}\aftertabspace
 \caption{Compression effectiveness for $\dtdw$-format word-level
indexes as a function of block size $B$, in bytes per posting.
Encoding parameter $F=3$ was used throughout.
All sizes include the hash table, terms and all other vocabulary
information, plus the postings themselves.
The best value in each row is shown in blue.
\label{tbl-blocksize-words}}
\end{table}

Word-level indexes contain more information and are thus more
expensive to store than document-level ones.
Table~\ref{tbl-blocksize-words} gives a full range of results, with
the listed bytes/posting rates again including the hash array, the
postings overhead, and all of the extensible list overheads, and
hence comparable with the values reported in
Table~\ref{tbl-blocksize}.
For word-based indexes employing the {\Const} strategy the best
compression effectiveness is achieved with somewhat larger block
sizes.
Note that in Table~\ref{tbl-blocksize} the compression rates are also
given in bytes per posting, but that there are more postings in a
word-level index than in a document-level index.
Table~\ref{tbl-datasets} gives exact ratios connecting these various
quantities.

The results in Table~\ref{tbl-blocksize-words} compare very favorably
with the $6$--$7$ bytes per posting (that is, per input word)
achieved by the best method of {\citet[Table 6, method
{\method{FibonacciB}}]{hb17adcs}} for the same mode of word-level
indexing.
They also compare well against the rates achieved by {\citet{bc05tr}}
for their schema independent word-level index (see the last row of
Table~\ref{tbl-indexes} in Section~\ref{sec-indexing}).
In particular, {\citeauthor{bc05tr}} report (in their Table~4) space
consumptions that equate to approximately $1.8$ bytes per posting
(method $\method{Explimit}_{16,1.1,512}$), but with that rate not
including the cost of the additional document-to-word mapping that is
required, nor the cost of the term vocabulary structure.
The schema-independent scheme is also unlikely to resolve queries as
quickly as our approach, because of the need to consult the document
mapping for each posting.
It thus represents a further option in the trade-off space
illustrated by Figure~\ref{fig-sausages}.

\subsection{Compared to Previous Immediate-Access Indexing Approaches}
\label{sec-likeforlike}

\begin{table}
\centering
\sisetup{
round-precision=0
}
\renewcommand{\tabcolsep}{1.0em}
\begin{tabular}{
l
	S[table-format=3.0]
		S[table-format=5.0]
}
\toprule
Approach
	& {Time (seconds)}
		& {Space (MiB)}
\\
\midrule
{\citet{hb17adcs}}
	& 329
		& 18114
\\
Ours
	& 545
		& 5603
\\
\bottomrule
\end{tabular}\aftertabspace

 \caption{Immediate-access indexing techniques, comparing
our implementation with that of {\citet{hb17adcs}}, building a
word-level index for the {\wiki} collection.
\label{tbl-lets-beat-dave}}
\end{table}

Table~\ref{tbl-lets-beat-dave} compares our new mechanism with the
code of {\citet{hb17adcs}}, building a $\dtdw$-style word-level
inverted index for the {\wiki} collection in each case.
The {\citeauthor{hb17adcs}} code executes more quickly than does
ours, but our software generates an index that requires only $30$\%
of the space.
That is, the careful attention we have given to minimizing the memory
footprint means that our code can process more than three times as
many documents in a given amount of main memory before needing to
transfer index data to secondary storage.

\subsection{Variable Block Indexing}
\label{sec-varblocks}

The mechanism described in Section~\ref{sec-somethingnew} uses
uniform-sized blocks, each of $B$ bytes, which is the $\Const_B$
approach of {\citet{bc05tr}} and {\citet{hb17adcs}} (noting that in
some descriptions $B$ is counted in postings rather than in bytes;
here we primarily make use of quantities measured in bytes).
The $\Const$ approach has a number of advantages, primarily the
simplicity of addressing, but has the disadvantage of being
non-adaptive to term frequencies -- rare terms are treated exactly
the same as common ones.
In particular, if $h$ bytes of linking (pointer) information are
required in each block, leaving $B-h$ bytes available for the payload
bytes storing vocabulary information and postings, then
constant-sized blocks result in a minimum {\emph{asymptotic overhead
ratio}} of $h/(B-h)$.
For example, if $B=64$ and $h=4$, the minimum asymptotic overhead
ratio of non-payload bytes to payload bytes is $\num{6.66666}$\%.
There will also be up to $B-1$ unused bytes in the last block in the
chain, which pushes the average overhead ratio higher.

A range of non-constant approaches were also explored by
{\citet{bc05tr}} and {\citet{hb17adcs}}, in particular, a range of
{\emph{exponential growth}} schemes, denoted $\Expon_{B,k}$.
In the $\Expon_{B,k}$ scheme $B$ is the first block size, and $k\ge1$
is a growth parameter.
The first block in each chain contains $B$ bytes in total, including
$h$ bytes of link overhead; the second contains a total of $\lceil
h+(B-h)k\rceil$ bytes; the third $\lceil h+(B-h)k^2\rceil$ bytes; and
so on, forming a geometric sequence of ever-increasing payload
capacities.
While this approach reduces the number of blocks in each chain from
being linear in the total payload volume to being logarithmic in the
payload volume, and thus similarly reduces the number of pointers
required, it doesn't reduce the average asymptotic overhead ratio.
In particular, as the blocks become larger, the average number of
unused bytes in the tail block also grows exponentially.
Asymptotically, if the block-on-block payload growth rate is by a
factor of exactly $k$, then the tail block contains a fraction of the
total payload capacity that can be calculated as
\begin{equation}
	\lim_{z \rightarrow \infty}
		\frac{B'k^{z}}{B' + B'k + B'k^2 + \cdots + B'k^z}
	= \lim_{z \rightarrow \infty}
		\frac{k^z}{(k^{z+1}-1)/(k-1)}
	\approx \frac{k-1}{k} \, ,
		\label{eqn-exponfrac}
\end{equation}
where $B'=B-h$ is the payload capacity of the first block, and where
$z$ is the index of the current tail block.
That is, in a postings list allocated with a current total capacity
of $n$ payload bytes, approximately $n(k-1)$ of those payload bytes
are in the tail block; moreover, when averaged across all postings
lists and as $k$ approaches $1$ from above, approximately half of the
index's tail block bytes will be unused at any given time.
For $k=1.1$, the growth rate favored by {\citet{bc05tr}}, that
implies an average asymptotic overhead ratio in the vicinity of
$5$\%, not including the approximately $\log_{k}(n/B)$ link pointers,
each consuming $h$ bytes.

That is, growing the blocks exponentially using any fixed radix $k$
incurs an overhead ratio that is again (in an amortized sense) linear
in the volume of stored payloads.
In both the {\Const} and {\Expon} cases the exact asymptotic rate is
determined by the parameter choice ($B$ and $k$, assuming $h$ is
fixed), and hence can be controlled to a certain extent.
But setting the parameters to minimize the asymptotic rate also has
the perverse effect of increasing the actual measured cost in typical
non-asymptotic (that is, practical) situations.
Care -- and a degree of empirical exploration to establish exact
overhead rates as percentages of the payloads stored
{\citep{bc05tr,hb17adcs}} -- is required if space is to be minimized.

Schemes that exponentially grow until some upper limit is reached and
then revert to constant block sizes {\cite{bc05tr}}, and schemes
based on Fibonacci numbers or that have a number of same-size repeat
blocks at each block size {\cite{hb17adcs}} all share this same
asymptotic behavior -- the expected overhead converges to a fixed
linear fraction of the total payload volume.

\subsection{Extensible Lists Revisited}
\label{sec-triangle}

To reduce the asymptotic overhead ratio a different approach is
required.
To set the scene, suppose fixed blocks of $B$ bytes each (including
the $h$ bytes required for the block's link pointer) will be
employed, and that a total payload of $n$ bytes is to be
accommodated, but with $n$ unknown.
Then the number of blocks required is $z = \lceil{n}/(B-h)\rceil$;
and the total overhead cost will be given by $W=Bz-n$, covering both
the links and the unused payload slots in the tail block.
If we assume that $n$ is large enough that $z \approx 0.5+n/(B-h)$
(because we don't actually know the value of $n$, but can assume that
when taken modulo $B-h$ it falls mid-way through the range from $0$
to $B-h-1$) then the wastage $W$ is
\[
	W = B\cdot \left(0.5+\frac{n}{B-h}\right) - n
	= 0.5\cdot B + n\cdot\frac{B}{B-h} - n\, .
\]
Making the further assumption that $h\ll B$, so that $B/(B-h) \approx
1+h/B$, we then get
\[
	W = 0.5\cdot B + \frac{hn}{B} + n - n
		= 0.5\cdot B + \frac{hn}{B} \,.
\]
Taking $h$ and $n$ to be fixed, and choosing $B$ to minimize $W$ then leads to:
\[
	\frac{\partial W}{\partial B}
	= 0.5 - hn/B^2
\]
which is equal to zero when
\[
	0.5 = hn/B^2
\]
that is, when
\begin{equation}
	B = \sqrt{2hn} \, .
	\label{eqn-definetriangle}
\end{equation}
As an example of what this means, consider the case when $h=4$ bytes
and a list has $n=20{,}000$ payload bytes.
Then a block size of $B = 400$ bytes is computed from
Equation~\ref{eqn-definetriangle}, and the resultant postings list
arrangement would have $z=\lceil 20{,}000/396 \rceil=51$ blocks, a
total of $20{,}196$ available payload bytes, a total of $204$ bytes
of link pointers, and a further $196$ bytes of unused space in the
tail block, for a total overhead of $400$ bytes.
Note how the overhead ratio is minimized when the overhead consists
of approximately equal volumes of link pointer bytes and unused tail
block bytes.

Of course, the problem is that we don't know the final value of $n$.
Indeed, there is no final value, $n$ is perpetually subject to upward
revision as the index is extended.
But $n$ does have a {\emph{current}} value, and we know exactly what
it is -- it is the sum of the payloads stored in the current chain of
blocks.
In particular, suppose that the $z$\,th block of each postings list
is of size $B_z$ and carries a payload of $p_z=B_z-h$ bytes.
Suppose further that the first $z$ blocks of the posting list are
full and hence that $n=\sum_{i=1}^z p_i$, and that we must determine
a block size for the $z+1$\,th block.
For the $\Const_B$ approach {\citep{bc05tr,hb17adcs}} the situation
is simple:
\begin{equation}
	B_{z+1} = B_{z} = B \, ,
	\label{eqn-constgrowth}
\end{equation}
as already noted in Section~\ref{sec-dynamic}.
The relationship captured in Equation~\ref{eqn-exponfrac} means that
the $\Expon_{B,k}$ method with growth parameter $k$ is also easy to
now state:
\begin{equation}
	B_{z+1} = h + (k-1) \cdot \sum_{i=1}^z p_{i} = h + (k-1) \cdot n\, .
	\label{eqn-expongrowth}
\end{equation}
In practice we may wish to use block sizes that are integer multiples
of the base unit $B$, and in this case we take $B_1=B$, and then
$B$-align each subsequent block $B_z$, taking care that the minimum
block size allocated is $B$:
\begin{equation}
	B_{z+1} = B \cdot
	\left\lceil
		\frac{h+(k-1)\cdot n}{B}
	\right\rceil \, .
	\label{eqn-exponrounded}
\end{equation}
For example, when $B=16$, $h=4$, and $k=1.5$, the sequence of
block sizes will be $B_z = \langle
16,16,16,32,48,64,96,144,208,\cdots\rangle$ bytes.

\begin{figure*}[t]
\centering
\includegraphics[width=0.6\textwidth]{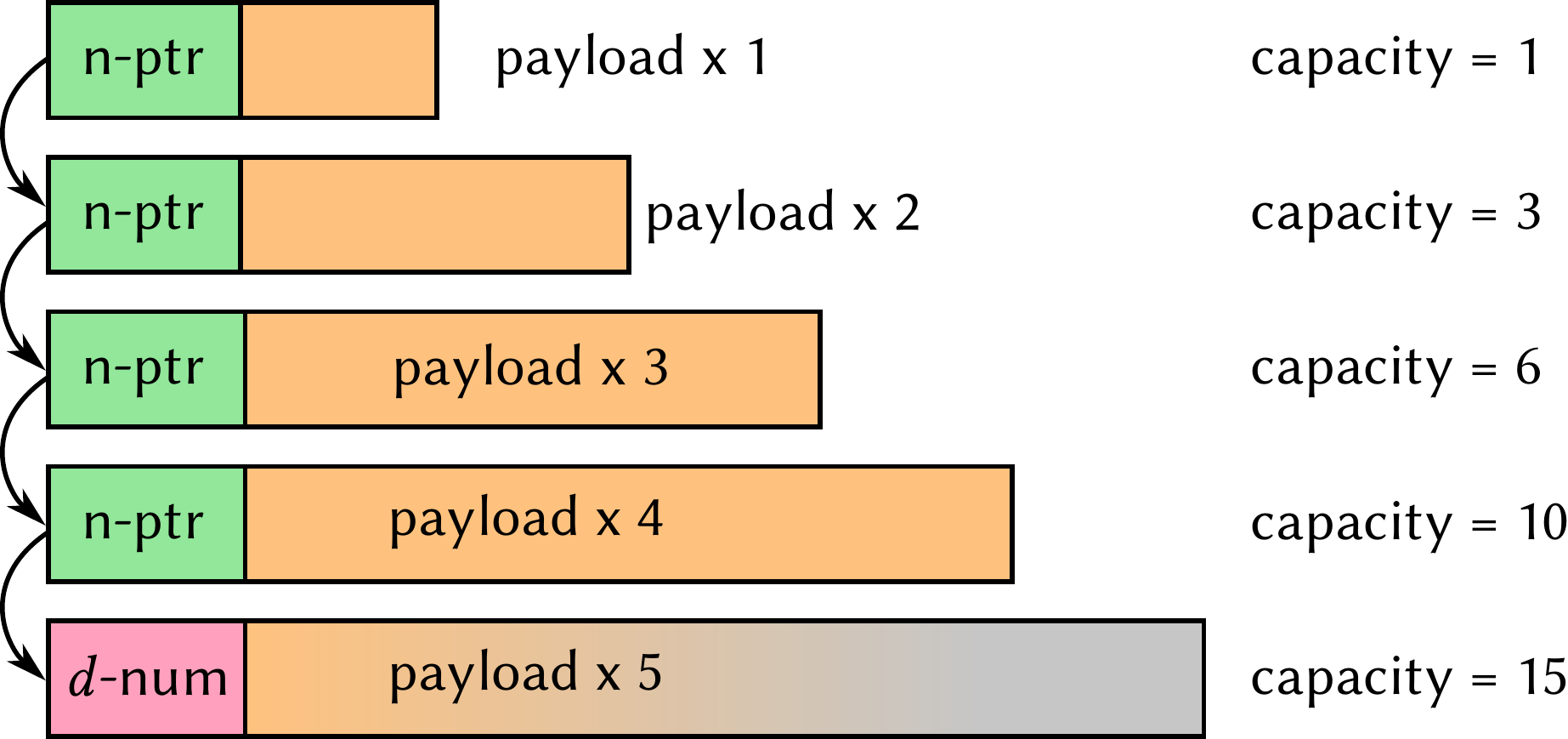}
  \caption{The basic {\Triangle} method for extensible lists,
starting with a single payload, and with each link pointer requiring
the same space as a single payload (that is, $B=2$ and $h=1$).
The $z$\,th block in the chain contains $z$ payloads, and the total
payload capacity in the first $z$ blocks is $n=z(z+1)/2$.
\label{fig-trian}}
\end{figure*}

Given the context established by the $\Const$ and $\Expon$ approaches
and their definitions via Equations~\ref{eqn-constgrowth}
and~\ref{eqn-exponrounded}, we now define a new growth mechanism for
extensible lists that exploits the relationship captured in
Equation~\ref{eqn-definetriangle}.
To understand the basis of this new {\Triangle} strategy, consider
Figure~\ref{fig-trian}, which shows a simple example that commences
with an initial block of size $B=2$ and requires $h=1$ cells for each
link pointer.
In the figure the $z$\,th block allocated always contains space for
$z$ payload slots, plus a link pointer.
Because $\sum_{i=1}^z i = z(z+1)/2 \approx z^2/2$, once $B_z$ is
allocated the structure contains approximately $z^2/2$ payload slots,
and $z$ links.
Moreover, the worst overhead ratios occur straight after each $z$\,th
block is added, when there is just one payload slot within it that
has been consumed.
At that moment the structure contains $z$ link pointers and
$z(z+1)/2$ payload slots, of which $n=(z-1)z/2+1$ are in use and
$z-1$ are vacant.
The overhead ratio at this point in the growth cycle is thus
\[
	\frac{2z-1}{n}
	=
	\frac{2z-1}{z^2/2 - z/2 + 1}
	\approx
	\frac{4}{z}
	\approx
	\frac{2\sqrt{2}}{\sqrt{n}}
\, .
\]
This is asymptotically superior to both the $\Const$ and $\Expon$
strategies, which have overhead ratios that are constant.
Note the close connection between this example and the conclusion
captured in Equation~\ref{eqn-definetriangle}.
In Figure~\ref{fig-trian} the relationship $n=z(z+1)/2$ means that
for a given value $n$, the largest block will be of size $B_z
\approx\sqrt{2n}$, and hence that the next block will be one larger:
$B_{z+1}\approx 1+\sqrt{2n}$, matching
Equation~\ref{eqn-definetriangle} in the $h=1$ case.

More generally, the $\Triangle_B$ method (named as a
consequence of the visualization shown in Figure~\ref{fig-trian} for
the simplest $B=2$ and $h=1$ case) is defined via:
\begin{equation}
	B_{z+1} = B \cdot
	\left\lceil
\frac{h + \sqrt{2hn}}{B}
	\right\rceil \, 
	\label{eqn-trianrounded}
\end{equation}
in which $n$ is again the total payload volume at the moment the
expansion is required.
With this formulation block sizes grow as the square root of what has
already been accommodated in the list, a more sedate pattern than the
rapid $\Expon$ acceleration.
For example, when $B=16$ and $h=4$, Equation~\ref{eqn-trianrounded}
leads to the sequence of block sizes $B_z=\langle
16,16,32,32,32,48,48,48,48,\cdots\rangle$, and hence to the sequence
of payload capacities $p_z=\langle
12,12,28,28,28,44,44,44,44,\cdots\rangle$.

\begin{figure}[t]
\centering
\includegraphics[width=0.7\textwidth]{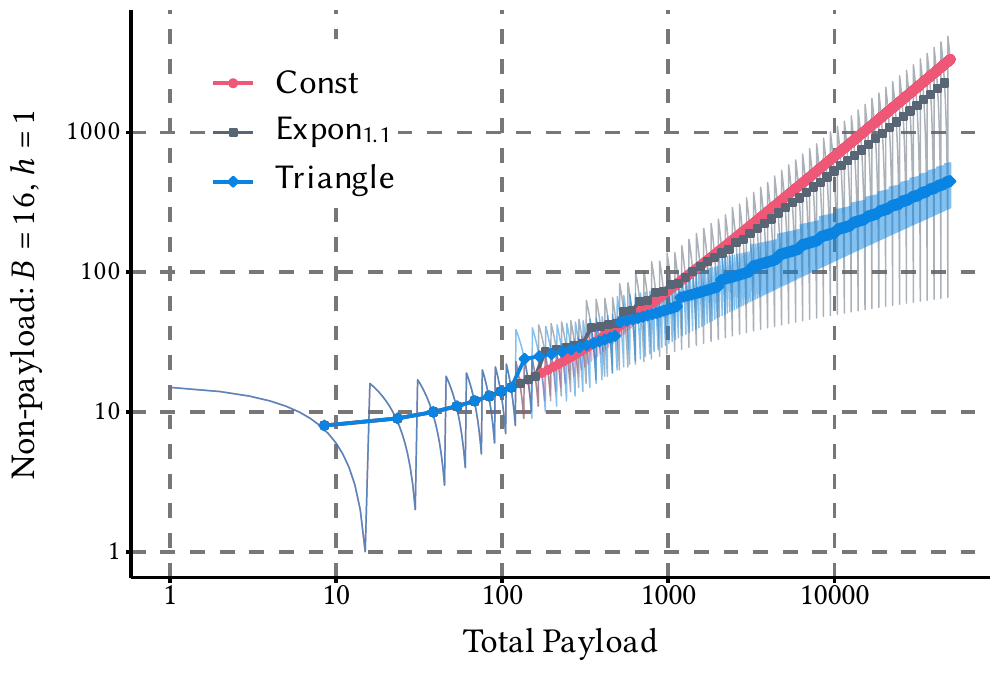} 
\caption{Volume of non-payload storage required, plotted as a
function of the total volume of payload data, where the first block
contains $B=16$ units of payload, and where $h=1$ element within each
block is required for the block link information.
The $\Expon_{16,1.1}$ and $\Triangle_{16}$ approaches are defined by
Equations~\ref{eqn-exponrounded} and~\ref{eqn-trianrounded}
respectively, with block sizes $B_z$ that are always integer multiples
of~$B$.
\label{fig-groblocks}}
\end{figure}

Figure~\ref{fig-groblocks} show the pattern of non-payload storage
required, as a function of the total volume of payloads, here with
first blocks of size $B=16$, and with $h=1$ cells required by each
link pointer and thus unavailable as payload (a scenario that
corresponds to $B=64$ and $h=4$ when the unit of measurement is
bytes).
The three sawtooth lines show the exact overhead count associated
with input payload volume for the three different growth methods,
with the sawtooth jumping to a new high point each time an additional
block is allocated, and then decreasing through a cycle as that block
gets filled.
The three smooth lines though the marked dots represent the average
overhead within each of the sawtooth growth cycles, and thus indicate
the long-term amortized overheads of the three methods.
The red ($\Const$) and grey ($\Expon_{16,1.1}$) lines become parallel
to each other in this log-log graph as the payload volume increases,
showing that they have the same asymptotic growth rate.

On the other hand, the overhead ratio for the new {\Triangle}
strategy is a sub-constant fraction of the number of payload words, a
relationship confirmed by the reduced gradient of the corresponding
blue line in the log-log graph.
Hence, while {\Triangle} is slightly less compact than both $\Const$
and $\Expon$ over small sections of the range of values $n$, it
always becomes more efficient than both $\Const$ and $\Expon$ on long
lists.
In Figure~\ref{fig-groblocks} the last full {\Triangle} growth cycle
covers lists containing between $n=49{,}681$ and $n=49{,}999$ payload
items inclusive, an average of $n=49{,}840.5$; and gives rise to an
average overhead ratio of just
$\num{0.898876}$\%
through that growth cycle.

\begin{table}[t]
\centering
\renewcommand{\tabcolsep}{1.0em}
\begin{tabular}{
c
	cS[table-format=1.3,round-precision=3]
		S[table-format=1.3,round-precision=3]
			cS[table-format=1.3,round-precision=3]
				S[table-format=1.3,round-precision=3]
}
\toprule
\multirow{2}{*}{Method}
	&& \multicolumn{2}{c}{Document-level}
			&& \multicolumn{2}{c}{Word-level}
\\
\cmidrule{3-4}\cmidrule{6-7}
	&& {$B=48$}
		& {$B=64$}
			&& {$B=48$}
				& {$B=64$}
\\
\midrule
{$\Const_B$}
	&& 2.09138
		& 2.10995
			&& 2.4484
				& 2.40379
\\
{$\Expon_{B,1.1}$}
	&& 1.99556
		& 2.05629
			&& 2.2628
				& 2.28233
\\
{$\Triangle_B$}
	&& \color{blue} 1.95172
		& \color{blue} 2.01265
			&& \color{blue} 2.22604
				& \color{blue} 2.24633
\\
\bottomrule
\end{tabular}\aftertabspace
 \caption{Total index cost (bytes per posting, comparable with
Tables~\ref{tbl-blocksize} and~\ref{tbl-blocksize-words}) for indexes
for {\wiki} using three different growth methods for extensible lists
and two different block sizes, both in conjunction with $h=4$.
In the $\Expon_{B,k}$ and $\Triangle_B$ methods the vocabulary in
each head block (see Figure~\ref{fig-layout}) requires two extra
bytes compared to $\Const_B$, to handle the variable block sizing.
The best value in each column is shown in blue.
\label{tbl-extensible}}
\end{table}

The critical question then is what happens in practice for a typical
mix of postings lists lengths.
If most lists are long, or if some lists are very long, {\Triangle}
is likely to lead to the most compact index.
On the other hand, if there are many lists that fall into the middle
band of lengths, then one of the other two methods might be the most
compact in a practical (rather than asymptotic) sense.
Another factor to be accounted for is increased complexity -- each
head block (see Figure~\ref{fig-layout}) needs an additional field
inserted, to track the current block size $B_z$ (or rather, to track
$z$); and the field $\var{nx}$ must also be extended, since $256$
bytes is no longer a plausible upper bound on block size.
In our implementation these small changes add two bytes to every head
node, with block sizes capped at $2^{16}$ bytes and $\var{nx}$
becoming a two-byte integer, and with $z$ a one-byte integer and
capped at $256$.

Table~\ref{tbl-extensible} shows compression effectiveness, returning
to $h=4$, with $B=48$ and $B=64$, with $n$ counted in bytes, with two
extra vocabulary bytes for the {\Expon} and {\Triangle} methods, and
with all other aspects as described in
Section~\ref{sec-somethingnew}.
The {\Triangle} approach secures small additional savings in all
cases compared to both the $\Const$ strategy and the $\Expon$
approach.

However, there is a drawback to both the {\Expon} and {\Triangle}
approaches, a result of their ever-lengthening postings blocks.
In conjunctive querying modes (and also in ranked querying modes when
implemented using a dynamic pruning mechanism such as {\maxscore} or
{\wand} {\citep{bc+03-cikm,dm+16-ipm,mm20cikm,tf95-ipm}}) the
$\var{seek\_GEQ}()$ operation plays an important role, bypassing
blocks that are not required.
But the likelihood of bypassing any particular block decreases as the
blocks become longer, because each block contains more postings.
That effect is then compounded, because more sequential decoding
effort is also required to reach any given document number within a
long block if it does get decoded.
In combination those two effects mean that while plain disjunctive
queries operate at similar speeds in all of {\Const}, {\Expon}, and
{\Triangle} approaches, conjunctive evaluation is fastest with the
{\Const} extensible list structure.
This tension provides yet another example of the space-speed-indexing
tradeoffs that were illustrated in Figure~\ref{fig-sausages}.

\subsection{Periodic Collation}
\label{sec-rearranging}

As described in Section~\ref{sec-somethingnew}, in the {\Const}
approach the postings list for each term $t$ consists of a chain of
blocks, all of some fixed length $B$, each separated from the other
blocks associated with $t$ by blocks allocated to other terms.
Access to each new block when following $t$'s postings chain might
thus result in a cache-miss (or equivalently, a pre-fetch failure),
since the intervals between $t$'s blocks are likely to be highly
variable.
This, in turn, means that the indexed sequential access associated
with the $\var{seek\_GEQ}()$ operations is relatively costly, since
only a few bytes of memory (a $b$-gap and then the $\var{n\_ptr}$)
are accessed from within each block, with another long memory jump
often immediately following.

To ameliorate that cost, the final enhancement we propose is that of
{\emph{collating}} the index blocks.
This is a very simple operation -- somewhat akin to the ``list
traversal'' phase discussed by {\citet{hb17adcs}} -- that can be
undertaken periodically, perhaps when indexing/querying load is low.
To get started, a copy of the hash array is made, with ingest
operations temporarily suspended and querying operations carried out
as usual via the old hash array.
Each non-empty element in that copied hash array, corresponding to
some term $t$, is then visited.
First, $t$'s head block is written from $\Index$ to a sequential disk
file, and then each of $t$'s other blocks is written, through to and
including the tail block; all the while replacing the $\var{n\_ptr}$
and $\var{t\_ptr}$ fields with revised values determined by a counter
of blocks written to disk.
No other alterations are made to any of the blocks, and they retain
all of their remaining components completely unchanged.
As terms are processed, the entries in the copied hash array are
updated with the new offsets of the corresponding head blocks in the
reordered on-disk version of the index.
This relatively lightweight process results in a ``collated'' version
of $\Index$ on the disk.
A brief pause in query processing is then required while a
single binary read operation brings the permuted index back into
memory, overwriting the exact same space within $\Index$.
At the same time the new hash array replaces the old one; once that
is done, document ingestion and querying can be resumed.

\begin{table}[t]
\centering
\renewcommand{\tabcolsep}{1.2em}
\sisetup{
group-separator = {,},
round-mode = places,
round-precision = 2,
table-format = 2.2,
}\begin{tabular}{l c
	S[table-format=1.2]
		S
			cS
				S[table-format=3.2]
}
\toprule
\multirow{2}{*}{Collection}
	&& \multicolumn{2}{c}{Conjunction}
		&& \multicolumn{2}{c}{Top-10 disjunction}
\\
\cmidrule{3-4}\cmidrule{6-7}
&& {Mean} & {$P_{95}$} && {Mean} & {$P_{95}$} \\
\midrule

{\Const}, interleaved
  &&  8.52642620 & 32.02605 && 90.28812198 &  455.96095 \\

{\Const}, collated
  && 4.08321065 &  13.93690 && 83.86623989 & 426.36235 \\

{\Triangle}, interleaved
  &&  31.4888
  	& 123.672 
  && 83.7783 
  	& 424.409
\\

{\Triangle}, collated
  && 31.0743
  	& 122.166
  && 82.6337
  	& 419.934
\\

\bottomrule
\end{tabular}\aftertabspace
 \caption{Query times for the {\wiki} collection for
two different querying (Section~\ref{sec-queryspeed}), with indexes
formed via the {\Const} and {\Triangle}
strategies with $B=64$ and $h=4$.
Each pair of rows shows the query time with postings blocks in
``arrival order'' in the index array $\Index$, and then after the
collation process.
All times are in milliseconds per query, and show the mean across the
query set (Table~\ref{tbl-queries}) and the $95$\,th percentile query
execution time across the query set.
\label{tbl-sortblock}}
\end{table}

The post-collation index still has the structure shown in
Figure~\ref{fig-layout}, and it is only the interleaving of the
blocks within $\Index$ that is affected by the collation process --
the index remains both queryable and extensible.
However in the collated index all of each term's blocks are stored
contiguously, a change that under some circumstances notably improves
query processing times.
Table~\ref{tbl-sortblock} supports that claim.
The query set used in Section~\ref{sec-queryspeed} was executed
against four indexes for the {\wiki} dataset.
The first index is as described in Section~\ref{sec-somethingnew} and
already measured in Section~\ref{sec-experiments}, and is constructed
using the {\Const} strategy.
The second is a collated version of that, still with all blocks the
same length, occupying $B=64$ bytes each.
The third of the three indexes is an uncollated index constructed
using the {\Triangle} strategy, which saves a small amount of space
compared to {\Const}; and the fourth index is a collated version of
the third, still using the {\Triangle} pattern of block lengths, but
now with each term's postings in a continuous section of $\Index$.

As can be seen, when the index in constructed using the {\Const}
approach, conjunctive querying time is reduced by a factor of up to
two in the collated index compared to the ``postings arrival order''
original interleaved index, with tail latency (the two $P_{95}$
columns) also notably improved.
On this {\wiki} collection, collation allows our system to outperform
the compact {\sf{PISA-Interp}} system which has a mean latency of
$5.74$ milliseconds per query (compared to the $4.08$ milliseconds
shown in Table~\ref{tbl-sortblock}).
Detailed profiling experiments showed that the majority of this
speedup was due to improved cache behavior, with around $66\%$ fewer
cache misses after collation.
There are also small gains possible in connection with disjunctive
queries.

On the other hand, when the index is constructed using the
{\Triangle} strategy, collation has little effect on execution times,
since the long postings lists are already represented as a small
number of long blocks.
Moreover, the long blocks of the {\Triangle} approach actively hinder
conjunctive query evaluation, an effect noted earlier in
Section~\ref{sec-triangle}.
With long postings blocks for common terms, skip-search using
$\var{seek\_GEQ}()$ operators is much less effective than it is with
the {\Const} strategy and its small blocks, suggesting that the
collated {\Const} mechanism provides the best overall balance of
features.

The collation process also has an associated cost.
On the {\wiki} collection, writing the complete index to SSD in a
single operation (that is, without collation) requires around $0.9$
seconds; the sequence of random accesses into $\Index$ and the large
number of small $B=64$-byte writes during collation extends that time
to $6.7$ seconds.
During that whole collation period ingestion must be stalled, meaning
that queries that arrive during that period will not be truly
immediate-access, with documents that arrived during the last
approximately $7.5$ seconds not retrievable.
If that risk is acceptable, then collation allows subsequent queries
to be resolved more quickly, thus providing yet another trade-off
option in the space shown schematically in Figure~\ref{fig-sausages}.
It may also be possible to handle tail blocks in a more strategic
manner and allow concurrent ingestion and collation, an option that
we plan to explore in future work.

\subsection{Managing Large Arrays}

A key element of our proposal is the use of a large
array of $B$-byte blocks.
If sufficient memory can be allocated to that array as a monolithic
segment, then all is well, and an immediate-index can be constructed
within that slab of memory.
On the other hand, if multiple indexes must be constructed at the
same time -- for example, if the arriving document stream contains
multiple languages, and each language is to be indexed independently
-- it may be desirable for the required arrays of blocks to somehow
co-exist within the same total amount of memory space, with each of
them growing as required until the overall envelope of space has been
exhausted.
In separate work we consider how to achieve that goal
{\citep{adcs22mm}}.

 \section{Discussion and Conclusions}
\label{sec-conclusions}

We have developed new insights into how dynamic retrieval systems can
be structured and organized.
Furthermore, as specific and concrete innovations, we have described
a byte-packing mechanism that greatly improves the compression
effectiveness of the previous {\vbyte} approach, and have also
developed a new extensible list technique that is asymptotically more
efficient than any of the mechanisms that have been considered
previously.
The second improvement reduces the overhead storage space to retain
$n$ items from being linear in $n$ (that is, $\Theta(n)$) to being
sub-linear in $n$ (to be precise, $\Theta(n^{1/2}) \in o(n)$).
Both of these enhancements represent significant developments in the
way that dynamic indexes are structured and stored.

As a broader assessment of what we have developed in this project, if
information retrieval was an athletics contest, then our structure
wouldn't be the fastest sprinter (because for sheer query speed, the
application of document reordering to a static document pool, with
variable length blocks, and {\maxscore}- or {\wand}-based dynamic
pruning would result in faster query processing); nor would it be the
highest jumper (because for minimal memory footprint, interpolative
coding and/or localized parameterized codes are better); and it might
not win the shot put either (because for sheer indexing speed, an
offline sort-based approach might be preferable).

But those ``better'' options -- occupying the corners in the triangle
in Figure~\ref{fig-sausages} -- do not offer dynamic indexing and
immediate-access querying, and hence cannot be used for
every component of a system of the type shown in
Figure~\ref{fig-mode}.
Given the desire to provide real-time support for document ingestion
and online search, the event (to continue the athletics metaphor) we
are competing in is the heptathlon, in which an all-round performer
must be competent across multiple dimensions, so as to achieve a
superior overall balance of performance.
In describing our fixed-block inverted file structure and the
immediate-access querying that it supports, and documenting its
performance relative to a carefully-engineered offline reference
system, we have explored the interior of the triangle in
Figure~\ref{fig-sausages}, and have presented an Olympic-level
heptathlete that will be of considerable interest to practitioners
and researchers alike.

\subsection{Limitations}

Our scheme does, of course, have limitations.
We have not considered document deletions nor updates, and at present
our software only handles ``append document'' and ``keyword search''
operations.
Deletions are a complex challenge in all inverted file-based
retrieval systems, because of the need to withdraw postings from the
middle of lists, and then adjust neighboring $d$-gaps; and the
immediate-index we have described is no different in that regard.
We have also only considered query processing in a ``proof of
concept'' manner in the experiments described in
Section~\ref{sec-experiments}.
For example, mechanisms of the type summarized in
Section~\ref{sec-pruning} that make use of stored block upper bounds
might allow faster similarity-based top-$k$ query processing, with
only small growth of the index being needed.

Nor have we considered the integration of the
immediate-access part of the index with the static shards.
As with all distributed systems, load balancing is important, and in
focusing entirely on the immediate-access index, we have deferred
consideration of how the shards shown in Figure~\ref{fig-mode} will
interact with each other.
In addition, we have not considered concurrency
control and the extent to which low-level operations require locking,
and have assumed in this discussion that all postings associated with
each ingested document are processed into the index before the next
query operation is permitted.
Finer-grained analysis of querying and document insertion, and of
in-parallel insertion of multiple documents, are also necessary in a
production system.

\subsection{Future Work}

Each of those limitations is also an opportunity for future work.
Our immediate next goal will be to consider how to integrate
responsive querying modes that -- with as little additional data
being stored as is possible -- allow more precise disjunctive
skipping to take place, so that we can reduce the time taken to carry
out ranked bag-of-words querying using similarity scoring models such
as {\method{BM25}}.
 
\myparagraph{Software}

Public software that implements our method is available from
{\url{https://github.com/JMMackenzie/immediate-access}}

\section*{Acknowledgements}

We thank the referees for their detailed and helpful input.
This work was in part supported by the Australian Research Council
(project DP200103136).

\end{document}